\documentclass[]{article}
\usepackage[english]{babel}
\usepackage[a4paper,top=2cm,bottom=2cm,left=3cm,right=3cm,marginparwidth=1.75cm]{geometry}

\usepackage{setspace}
\usepackage[inkscapeformat=png]{svg}
\usepackage{import}
\usepackage{siunitx}
\usepackage{placeins}
\usepackage{enumitem}
\usepackage{amsmath}
\usepackage{amsfonts}
\usepackage{graphicx}
\usepackage[colorlinks=true, allcolors=blue]{hyperref}
\usepackage{caption}
\usepackage{subcaption}
\usepackage{yfonts}
\usepackage{cite}
\usepackage{colortbl} % Required for coloring cells
\usepackage{array}    % Required for centering in the m column type
\usepackage{multirow}
\usepackage{comment}
\begin{document}
\title{Acoustic wave scattering by spatio-temporal interfaces}
\def\affil#1{\begin{itemize} \item[] #1 \end{itemize}}

\author{\normalsize \bfseries J. Galiana , J. Redondo, R. Picó and V. J. Sánchez-Morcillo}
\date{}

\maketitle
%%% Otros títulos: 

% Dynamic Scattering Regimes in Moving Acoustic Interfaces
% Analytical and Numerical Analysis of Acoustic Scattering by Moving Boundaries
% Wave Scattering in Motion: Acoustics of Moving Slabs

\affil{\begin{center}\normalsize \textit{Instituto de Investigación para la Gestión Integrada de Zonas Costeras (IGIC), \\ Universitat Politècnica de València, Paranimf 1, 46730 Gandia, Spain}
\end{center}}

\begin{abstract}

Space-time materials are obtained by modulating a physical medium with a traveling-wave perturbation of one or several of its constitutive parameters, such as the density or the bulk modulus in the case of acoustic materials. When this modulation has the form of a moving and abrupt (subwavelength) transition between two parameter values, we refer to a spatio-temporal interface, which may be considered as a building block for more complex  space-time materials. This work considers the problem interaction and scattering of acoustic waves with a single spatio-temporal interface, and a sequence of two interfaces forming a slab.  Several regimes defined by the relation between the sound propagation velocities and the interface velocity (namely subsonic, intersonic, and supersonic regimes) are discussed. Analytical expressions for the frequency conversions and scattering coefficients are obtained, and compared with numerical simulations based on an equivalent FTFD squeme. 
\end{abstract}

%\tableofcontents

\section{Introduction}

Some physical problems, studied decades ago, are often revisited from a new perspective, motivated by advances in knowledge or improvements in analytical, numerical, or experimental techniques. This is the case for wave propagation in space-time-dependent media, which has become accessible to experimentation mainly due to progress in smart materials and metamaterials. This broad topic includes, as a particular case, the scattering of waves by moving interfaces, which is the subject of this work. Some well-known phenomena, such as the Doppler effect, are related to this problem, but the implications extend beyond frequency conversions and enable new ways of wave manipulation and control.  

The first studies on the reflection and transmission of sound by a moving medium date back to the 1950s \cite{Keller55,Franken56,Miles57,Ribner57}. The medium consisted of a layered fluid, with the layers in relative motion, separated by a flat interface. The fluids were assumed to flow parallel to the interface, so the interface itself was stationary. These studies already revealed new phenomena, such as changes in the spectrum of the scattered waves and velocity-dependent total reflection, among others.

After these seminal acoustical studies, most of the progress in this topic was achieved for electromagnetic fields in dielectrics. A space-time dependent wave velocity was introduced \cite{Landauer63}, invoking the existence of "materials whose wave propagation properties are subject to some degree of external control", a concept that anticipates the principles underlying modern metamaterials. Frequency changes in the scattered waves were also reported. At the same time, the dispersion relations for electromagnetic waves propagating in space-time periodic media were derived \cite{Cassedy63}. These works motivated further theoretical developments in the following years.

Wave interaction with moving dielectric interfaces was soon after considered \cite{Yeh65}, where scattering coefficients for reflected and transmitted waves were also obtained. The results were later extended to moving dielectric slabs \cite{Yeh66}. Similar results were obtained  by alternative methods\cite{Ramasastry67}. In the same year, a detailed study of the moving dielectric interface was presented \cite{Tsai67} and extended to the case of multiple moving interfaces (yet in a stationary medium), introducing the idea of periodically modulated space-time media, which is now experiencing a revival with the concept of space-time crystals. In all cases, the interface was modeled as an abrupt or step-like change in parameters. Later, these results were extended \cite{Landauer73} to consider a smooth modulation of the interface and also discussed the effect on nonlinearities in the propagating media. 
The acoustic problem was revisited later \cite{Kong70}, addressing the interaction of a plane incident acoustic wave with an $n$-layer stratified moving medium, where the flows are parallel to the interface. The differential equations governing wave behavior in such a moving medium were derived using first-order Lorentz transformations instead of Galilean transformations.  The more complex case of moving layers with a space-dependent velocity profile \cite{Steinmetz72} (such as a Poiseuille flow) was treated as a sequence of sublayers, each with constant a velocity, in a manner resembling the Riemann's integral concept.

Next, significant progress was made in a series of works by L.A. Ostrovskii \cite{Ostrovskii67,Ostrovskii71,Ostrovskii76}. The concept of a "moving parameter jump" introduced for the electromagnetic problem \cite{Ostrovskii67} is actually similar to the idea of a spatio-temporal interface presented here for acoustic waves.  Different cases were discussed \cite{Ostrovskii67}, including interfaces moving faster than light in the medium, to one or both sides of the jump. In this way, subluminal, superluminal, and interluminal situations are considered for the first time \cite{Ostrovskii71}.

All these studies have guided and inspired recent research in the field. After many years without substantial progress, a new stage in the study of space-time media began with the emergence of smart materials in acoustics and other areas. Electrostrictive and magnetostrictive solids, electro- and magneto-rheological fluids, and shape-memory materials form the basis for new materials with high manipulation capability. In this new context, the work on space-time (dynamic) materials by K.A. Lurie \cite{Lurie17}, who introduced concepts such as activated materials, is also relevant. Attempts to realize space-time interfaces are reported here as well.

Elastic waves in media with space-time modulation of some parameter have been also modelled \cite{Shui14}, and experimentally demonstrated \cite{Delory24}. Again for electromagnetic waves, important progress in this topic is reported in a series of works \cite{Deck-Leger19a,Deck-Leger19b}, and subsequent reviews \cite{Caloz20a, Caloz20b}. These studies are based on classical, established results and also introduce new concepts, such as space-time crystals.

This work explores the interaction of sound waves with moving interfaces and slabs, separating standing fluid media, by analyzing in detail the different regimes that occur when the interface velocity takes specific values relative to the sound speeds (namely, subsonic, supersonic, and intersonic regimes). The amplitudes of the scattered waves and the frequency and wavelength conversions are obtained analytically. The results are consistent with numerical modeling, which uses an approach based on a Centered-in-Time Finite-Difference Time-Domain (CIT-FDTD) scheme.

\section{Acoustic waves in moving media and moving interfaces}

\subsection{Model equations}

The propagation of small-amplitude sound waves in a --in general-- inhomogeneous and non-stationary lossless medium, defined by its ambient mass density $\rho$ and bulk modulus $\kappa$, is described by the equations of conservation of mass and momentum.
The form of the equations differ depending on the relative motion between the medium and the observer (here called the listener) \cite{Kong70}. If the medium is at rest with respect to a fixed reference frame (for example, a stationary listener), these equations are
\begin{subequations}
\label{eq:acoust1}
\begin{align}
&\rho\frac{\partial \mathbf{u}}{\partial t}+\nabla p = 0 ,\\
&\frac{\partial p}{\partial t} + \rho c^2 \nabla \cdot \mathbf{u} = 0
\end{align}
\end{subequations}

\noindent where $\mathbf{u}$ is the particle velocity vector, $p$ is the acoustic pressure, and $c=\sqrt{\kappa/\rho}$ is the speed of sound in the medium. These equations also describe the propagation of acoustic waves in a medium moving with respect to an observer who moves at the same velocity (i.e., without relative motion).

This set of equations can be easily extended to describe wave propagation in a medium moving uniformly at velocity $\mathbf{v}$ (representing a background flow) with respect to a listener at rest. The equations then take the form
\begin{subequations}
\label{eq:acoust2}
\begin{align}
\frac{\partial \mathbf{u}}{\partial t} + (\mathbf{v} \cdot \nabla)\mathbf{u} +  \frac{1}{\rho} \nabla p = 0 ,\\
\frac{\partial p}{\partial t} + \left( \mathbf{v} \cdot \nabla \right) p +\rho c^2  \nabla \cdot \mathbf{u} = 0 \end{align}
\end{subequations}
The same equations describe waves propagating in a static medium, as registered from a listener moving at the same velocity $\mathbf{v}$. Both descriptions are related by a Galilean change of coordinates: $\mathbf{r}'=\mathbf{r}+\mathbf{v}t,t' = t$. Equations (\ref{eq:acoust1}) and (\ref{eq:acoust2}) form the basis for the theoretical modeling of many acousticproblems, including those involving space-time modulations.

Note that the situation is different for electromagnetic waves, where the equations appear identical regardless of the observer's state of motion (after Lorentz transformations). As a consequence of this invariance, the measured electromagnetic fields depend on the velocity of the moving frame. This does not occur in acoustics, where the measured fields appear the same in different frames under Galilean or first-order Lorentz transformations \cite{Kong70}.

If we consider solutions in the form of plane waves,
\begin{equation}
\left( \mathbf{u},p\right) =\left( \mathbf{U},P\right)  e^{i(\omega t-\mathbf{k}\cdot\mathbf{r})}, 
\label{eq:pw1}
\end{equation}
the dispersion relation $\omega(\mathbf{k})$ is obtained by substituting (\ref{eq:pw1}) into the equations of motion. For the stationary case, the usual relation $\omega = c \left|\mathbf{k}\right| $ is obtained. However, for the moving case described by Eqs. (\ref{eq:acoust2}), dispersion relation results
\begin{equation}
\left( \omega-\mathbf{v} \cdot \mathbf{k} \right)^2 = \left( c \mathbf{k} \right)^2,
\label{eq:dispersion}
\end{equation}
It is apparent from Eq. (\ref{eq:dispersion}) that motion introduces a kind of anisotropy in the medium, which results in an asymmetry (direction-dependent) in the propagation of acoustic waves, which does not occur in a stationary medium.

In the following, we restrict for simplicity the analysis where wave-interface interactions take place only along $x$-direction; therefore, we assume plane waves at normal incidence. The analysis could be generalized to oblique incidence, but this is beyond the scope of this paper. 

\subsection{Spatio-temporal interface}
\label{sec:s_t_interface}

An acoustic interface is a boundary that separates two regions of space with different properties, such as two media (fluid or solid), each defined by its constitutive parameters. In linear, lossless acoustics, these parameters are the mass density $\rho$ and the bulk modulus $\kappa$, while in optics (or electromagnetism in general), they are the permittivity $\varepsilon$ and the permeability $\mu$. In both cases, the constitutive parameters determine the propagation speed of the wave and the specific impedance of the medium. We assume that the change in parameters from one medium to another is abrupt (step-like), which in practice means that the transition between the two media occurs over a spatial scale much smaller than the characteristic wavelength of the waves involved. If the interface moves at a constant speed $\mathbf{v}$, the parameters of the medium will also change locally in time, over a characteristic time much shorter than the period of the wave.  

\begin{comment}
Note, however, that the abruptness of the transition does not need to be simultaneous in space and time: an interface may be spatially sharp while evolving slowly in time, or it may change abruptly in time across a broad spatial region.
\end{comment}

The spatio-temporal dependence may result from two different situations, which lead to different types of problems: those involving moving media (including flows) and those involving moving boundaries (interfaces) between two standing media (also called a moving property interface\cite{Shui14}). These two problems are distinct but share important similarities, as first pointed out in \cite{Tsai67}. For example, a space-time interface can be created by a moving external action that alters the properties of a medium as it passes (such as its density and/or elasticity), producing the analogue of a moving wall, but with the material on each side of the interface remaining at rest.

A moving interface can be described as an inhomogeneous medium with a space-time dependent sound propagation velocity, given by
\begin{equation}
c(x,t)=c_1+(c_2-c_1) \theta(x-vt),
\label{eq:cxt}
\end{equation}
where $c_1$ and $c_2$ are the sound speeds on either side of the interface, and $\theta$ is the Heaviside function. Here the parameter $v$ represents the velocity of the moving interface. In contrast, a moving inhomogeneous medium would be described by a velocity $c'(x,t) = c(x,t) + v$, with $c(x,t)$ given by Eq. (\ref{eq:cxt}), which is actually a different problem. Now, the same symbol $v$ refers instead to the velocity of the material flow.

It is interesting to compare the problem of a moving interface separating two standing media, as in Eq. (\ref{eq:cxt}), with the problem of a standing interface in the presence of a constant flow on both sides of the interface. In this case, the velocity is described by
\begin{equation}
    c''(x,t)=v+c_1+(c_2-c_1) \theta(x),
    \label{eq:c2xt}
\end{equation}
Such a situation requires the interface to be transparent to the flow but not to the waves. Note that these two problems (a moving interface in a resting medium and a stationary interface with flow) are equivalent after a Galilean change of coordinates in Eq. (\ref{eq:cxt}), $x \rightarrow x+ vt$, and $c_i \rightarrow c_i + v$, with $i = 1, 2$. This equivalence is relevant for the numerical modeling proposed to test the validity of the theoretical results, as discussed in section \ref{sec:numerical}.                                                                        
To conclude this section, we recall that a moving interface presents two limiting cases: a stationary interface ($v = 0$) and a step-like temporal interface ($v \to \infty$). The first case corresponds to a spatial interface, where it is well known that the frequencies remain unchanged after reflection and transmission  (what implies energy conservation), while the wavelength of the transmitted wave changes due to the different propagation velocity in the second medium.

The second case corresponds to a purely temporal interface, representing a medium (the initial or early medium) whose properties change instantaneously at a given time throughout the entire spatial domain, with the new properties defining the second (also called \textit{later}) medium. The wave-interface interaction generates forward and backward traveling waves, each propagating at the speed of the second medium in opposite directions. In this case, the frequency changes, but the wavelength does not (momentum is conserved), after the interaction with the interface.

In the general case of a spatio-temporal interface considered here, both the frequency and the wavelength are modified after the interaction. In the next sections, we discuss in detail the interaction of acoustic waves with a single interface (section \ref{sec:interface}) and with a slab formed by two parallel interfaces moving at the same speed (section \ref{sec:slab}).  
%%%%%%%%%%%%%%%%%%%%%%%%%%%%%%%
\section{Wave interactions with a moving interface}
\label{sec:interface}

The interaction of an incoming wave with a moving interface, as previously stated, results in various scattered waves ($s$), which can be classified as incident $(i)$, reflected $(r)$, forward $(f)$, or backward $(b)$ waves. Each wave component on either side of the boundary can be represented as a plane wave in the form
\begin{equation}
\left( \mathbf{u}_s, p_s \right) =\left( \mathbf{U}_s, P_s \right) e^{i(\omega_s t \pm k_sx )},
\label{eq:sols}
\end{equation}
where $x$ is the coordinate normal to the interface. The amplitudes, frequencies, and wavelengths of each wave are generally different and can be determined by properly applying the boundary conditions at the moving interface and requiring the continuity of the phase at the boundary. 

Since three different velocities are involved in this problem (\(c_1\), \(c_2\), and \(v\)), several interaction regimes are expected, depending on their relative values. We discuss the different regimes separately.

\subsection{Subsonic regime}

\begin{figure}[h!]
    \centering
    \includegraphics[width=0.7\textwidth]{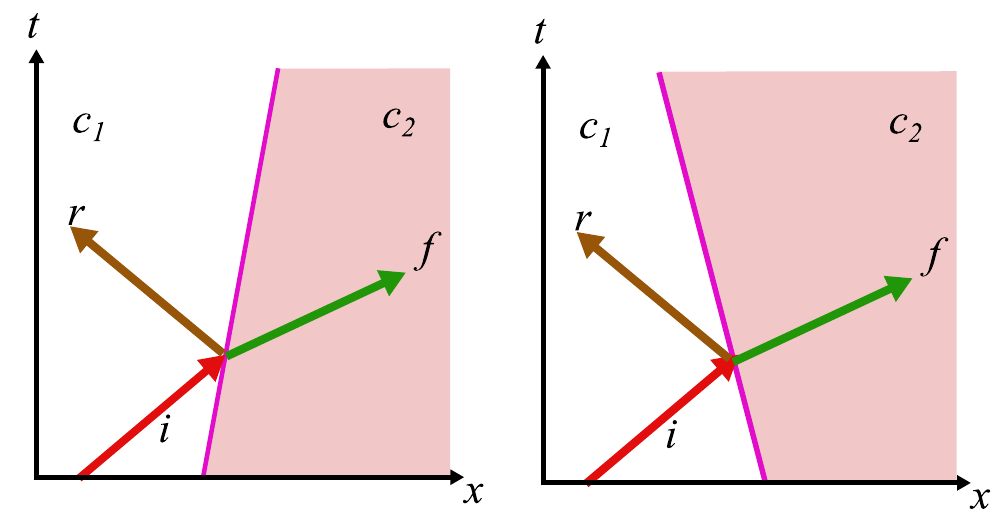}
    \caption{Interaction in the subsonic case, with a co-propagating (left) and counter-propagating (right) interfaces.}
    \label{fig:subsonic}
\end{figure}

When the interface speed is slower than the wave speeds in each medium, i.e. \( |v| < \min (c_1,c_2) \), the interaction regime is subsonic. It corresponds to the cases shown in Figs.  \ref{fig:subsonic}(a) and \ref{fig:subsonic}(b).  In this case, the incoming wave generates two scattered waves, one that propagates into the second medium in the same direction as the incident wave, called the \textit{forward} wave, and one that propagates in the opposite direction and remains in the first medium, called the \textit{reflected} wave. The stationary interface is a particular case of this regime \cite{kinsler}. 

Incident and reflected waves propagate in the first medium, in opposite directions, at speed $c_1$, while a transmitted wave propagates forward in the second medium at speed $c_2$. The following continuity conditions must be fulfilled at the interface for pressure and particle velocity, respectively 
\begin{align}
\label{eq:bcsub1}
(p_i+p_r)|_{x^{-}}&=p_f|_{x^{+}},\\ 
 \label{eq:bcsub2}
(v_i+v_r)|_{x^{-}}&=v_f|_{x^{+}},
 \end{align} 
where $x^{+}$ and $x^{-}$ denote the coordinate immediately at the right and at the left of the interface respectively, or  $x^\pm=x\pm \varepsilon$ with $\varepsilon\rightarrow 0$. For the uniformly moving interface, the coordinate changes with time as $x(t)=v t$ . Using the continuity of pressure in Eq. (\ref{eq:bcsub1}), and the relations between frequencies and wavenumbers in each medium: $\omega_i=c_1 k_i$, $\omega_r=c_1 k_r$, $\omega_f=c_2 k_f$, we can write the pressure continuity condition at the moving boundary as
\begin{equation}
P_i  e^{i\omega_i (1-\frac{u}{c_1})t}+P_r  e^{i\omega_r (1+\frac{u}{c_1})t}=P_f  e^{i\omega_f (1-\frac{u}{c_2})t} 
\label{eq:bcsub3}
\end{equation}
The validity of Eq.  (\ref{eq:bcsub3}), implies the equality of the phase factors. This leads to the values of the frequencies of reflected and transmitted waves relative to the incidence frequency, which can be expressed as  
\begin{eqnarray}
\Omega_r=\frac{\omega_r}{\omega_i}=\frac{1-\frac{v}{c_1}}{1+\frac{v}{c_1}}, \\
\Omega_f=\frac{\omega_f}{\omega_i}=\frac{1-\frac{v}{c_1}}{1-\frac{v}{c_2}}.
 \label{eq:freqs}
\end{eqnarray}
Note that, through the frequency–wavenumber relation, the constancy of sound speed within each medium implies a change in wavenumber when the frequency changes. The same relations for frequency change are also found in the electromagnetic problem. 

To find the scattering coefficients for reflected and forward (transmitted) waves, we first express velocities in terms of pressures, using Eq. (\ref{eq:acoust1}) and assuming solutions in the form of plane waves. From the relation $i \omega  \rho\mathbf{v}=-\nabla p=-i \mathbf{k}p$, we may write for the particular waves the relations
\begin{eqnarray}
  v_i=-\frac{p_i}{z_1},  v_r=\frac{p_r}{ z_1},  v_f=-\frac{p_f}{z_2}
  \label{eq:velocs2}
\end{eqnarray}
where $z_i=c_i \rho_i$ is the specific impedance of medium $i$.

The pressure scattering coefficients of the reflected and forward waves are defined as $\mathcal{R} = P_r / P_i$, $\mathcal{F} = P_f / P_i$. They can be obtained by solving Eqs. (\ref{eq:bcsub1}) and (\ref{eq:bcsub2}) and using the pressure-velocity relations in Eqs. (\ref{eq:velocs2}), yielding
\begin{eqnarray}
\mathcal{R}=\frac{z_2 - z_1}{z_1 + z_2}, 
\mathcal{F}=\frac{2z_2}{z_1 + z_2}
\label{eq:coefs1}
\end{eqnarray}
which are the same coefficients as for the static case ($v = 0$) \cite{kinsler}.

We note that scattering coefficients differ when evaluated for pressures or for particle velocities. Since for a given medium $p=\pm z v$, it is easy to show that $\mathcal{R}_v=-\mathcal{R}_p$, and $\mathcal{F}_v = (z_2 / z_1) \mathcal{F}_p$. In the following, coefficients without subindex refer to pressure waves.

\subsection{Supersonic regime}

\begin{figure}[h!]
    \centering
    \includegraphics[width=0.7\textwidth]{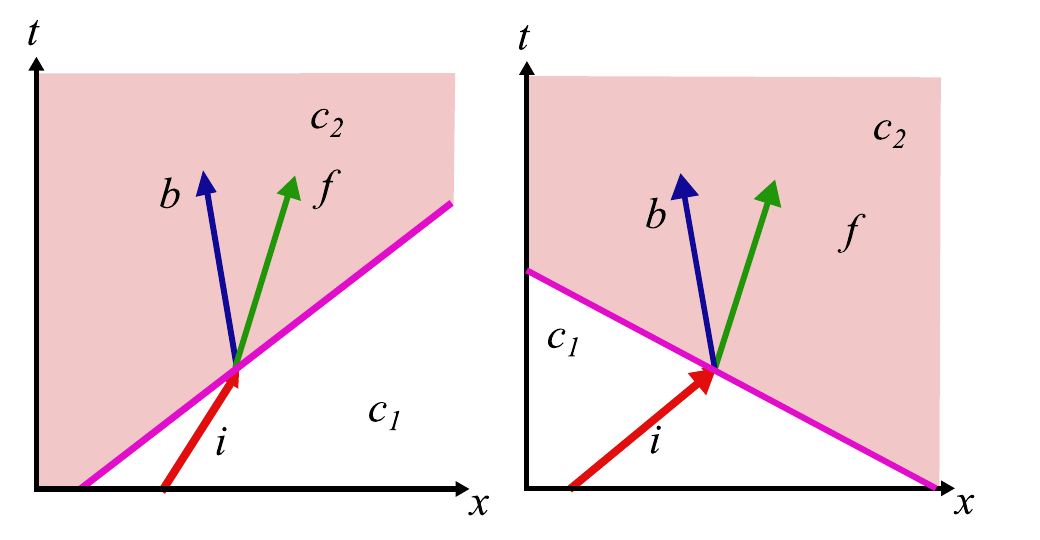}
    \caption{Interaction in the supersonic case, with a co-propagating (left) and a counter-propagating (right) interface.}
    \label{fig:supersonic}
\end{figure}

In this case, the interface speed is higher than the wave speed in both media, i.e., \( |v| < \max (c_1, c_2) \). This corresponds to the cases shown in Figs. \ref{fig:supersonic}(a) and \ref{fig:supersonic}(b). In Fig. \ref{fig:supersonic}(a), the interface moves in the same direction as the incoming wave (co-propagating case), and wave-interface interaction is possible only if the interface overtakes the incident wave. In both cases, the interaction again leads to the generation of two scattered waves, but in this case, both propagate within the second medium. These waves are called \textit{forward} and \textit{backward}, depending on their respective directions of propagation. Note that there is no reflected wave in the sense of the subsonic regime (i.e., no wave is scattered into the incident medium).

In the supersonic regime, the fields obey a different set of continuity conditions,
\begin{align}
 \label{eq:bcsup1}
 p_i|_{x^{-}} &= (p_b + p_f)|_{x^{+}}, \\
 \label{eq:bcsup2}
 v_i|_{x^{-}} &= (v_b + v_f)|_{x^{+}}.
\end{align}
Using the continuity of pressure in Eq. (\ref{eq:bcsup1}), and noting that now $\omega_{b} = c_2 k_{b}$, we may write the continuity of pressure at the boundary as
\begin{equation}
P_i  e^{i\omega_i (1-\frac{v}{c_1})t}=P_f  e^{i\omega_f(1-\frac{v}{c_2})t}+P_b  e^{i\omega_b (1+\frac{v}{c_2})t}, 
\label{eq:sols3}
\end{equation}
which, again, implies that the phase factors must be equal. This condition determines the frequencies and wavenumbers of the forward and backward waves. The frequency change for the forward wave is the same as in the subsonic case, while the frequency change for the backward wave is given by 
\begin{eqnarray}
\Omega_b=\frac{\omega_b}{\omega_i}=\frac{1-\frac{v}{c_1}}{1+\frac{v}{c_2}}.
\label{eq:freqs2}
\end{eqnarray}
The pressure scattering coefficients for the backward and forward waves are defined as $\mathcal{B} = P_b / P_i$, $\mathcal{F} = P_f / P_i$ . They can be obtained by solving Eqs. (\ref{eq:bcsup1}) and (\ref{eq:bcsup2}) and using the pressure-velocity relations in Eq. (\ref{eq:velocs2}), together with $v_b = z_2 p_b$. This yields
\begin{eqnarray}
\mathcal{B}=\frac{z_1-z_2}{2 z_1},  
\mathcal{F}=\frac{z_1+z_2}{2 z_1}.  
 \label{eq:coefs2}
\end{eqnarray} 
\subsection{Intersonic regime}

This regime occurs when the interface moves at a speed between the wave propagation speeds in the two media. As first noted by Ostrovskii,\cite{Ostrovskii67} this case requires special treatment because the continuity conditions at the boundary alone are insufficient to determine the amplitudes of the scattered waves, unlike in the subsonic and supersonic cases. The reason is that the number of scattered waves is either greater or fewer than the number of continuity conditions. Therefore, intersonic problems are actually under-determined or over-determined, respectively, since they involve three or one unknowns (the scattered amplitudes) for two continuity conditions.

\begin{figure}[h!]
    \centering
    \includegraphics[width=0.7\textwidth]{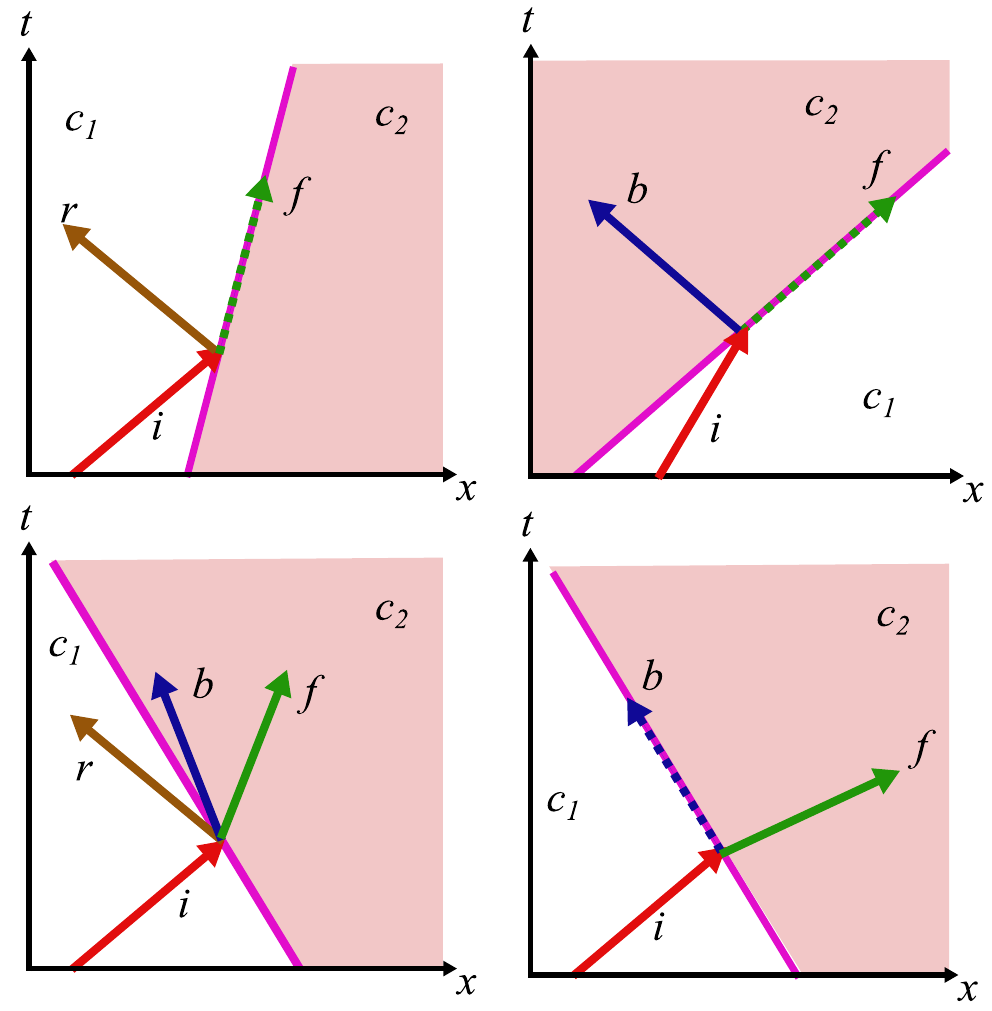}
    \caption{The four cases of intersonic interactions. Left colum shows the cases where $c_1>c_2$, in the co-propagating (top) and counter-propagating (bottom) cases. Right colum shows the case where $c_1<c_2$. }
    \label{fig:intersonic}
\end{figure}

Four combinations of velocities lead to this regime: \( c_1 < |v| < c_2 \) and \( c_2 < |v| < c_1 \), each resulting in a different scattering scenario. Fig. \ref{fig:intersonic} illustrates the different cases.

Only in one of the cases shown, corresponding to the bottom-left frame, the interaction results in three scattered waves. In the remaining situations, only one scattered wave results from the interaction (of type $r$, $b$, or $f$ in each case), accompanied by a shock wave attached to the interface, denoted by a dotted arrow \cite{Shui14}.

In \cite{Ostrovskii67}, a method for evaluating the scattering coefficients was proposed for the case of three scattered waves, \( c_1 > -v > c_2 \): the interface moving at speed $v$ is replaced by an auxiliary layer of thickness $\delta$, composed of a material with graded impedance and average value $z_m = \rho_m c_m$, varying from $z_2' = z_m + \Delta z$ to $z_1' = z_m + \Delta z$. The average impedance is chosen so that $c_m = v$, which is a synchronism point.

Then, the problem of a single interface is divided into two consecutive interfaces, one subsonic and one supersonic, where the previously obtained scattering coefficients can be used recursively. Finally, the original interface is recovered by taking the limit $\delta \rightarrow 0$. In this way, we obtain
\begin{eqnarray}
\mathcal{R}=\frac{z_m-z_1}{z_m+ z_1},  
\mathcal{B}=\frac{z_m-z_2}{z_m+ z_1},  
\mathcal{F}=\frac{z_m+z_2}{z_m+ z_1}  
 \label{eq:coefs3}
\end{eqnarray}
where $z_m=\rho_m c_m$ is the specific acoustic impedance evaluated at the synchronism point $c_m = v$. Note that this implies a dependence of the scattering coefficients on the interface speed $v$, unlike the subsonic and supersonic cases. There is still one unknown, $z_m$, because the acoustic properties of the auxiliary medium have not been fully specified. An expression depending only on the (known) velocities can be obtained if one of the medium parameters (density or bulk modulus) is assumed to be constant. In the case of constant density, $\rho_1=\rho_2=\rho_m$, we arrive at
\begin{eqnarray}
\mathcal{R}=\frac{v-c_1}{v+ c_1},  
\mathcal{B}=\frac{v-c_2}{v+ c_1},  
\mathcal{F}=\frac{v+c_2}{v+ c_1}  
 \label{eq:coefs3}
\end{eqnarray}
In the case where $\kappa$ is constant and $\rho$ is variable, the coefficients $\mathcal{R}$ and $\mathcal{B}$ change sign (recall that $z=\rho c=\kappa / c$, and $c=\sqrt{\kappa/\rho}$), while $\mathcal{F}$ remains unchanged.

Note that these scattering coefficients, unlike in the subsonic and supersonic cases, depend on the interface speed $v$.

\subsection{Interaction description in a reduced parameter space}
\label{section}

The different interaction regimes can be described in a simpler way in a reduced parameter space using only two parameters. This dimensionality reduction is achieved by taking the speed of the first medium, \(c_1\), as the reference for all velocities.

We define \(\gamma = c_2 / c_1 > 0\) as the ratio of the propagation velocities in the two media separated by the interface, referred to as the first and second medium. We recall that first medium refers to the medium in which the incident wave propagates before interacting with the moving interface. We also define the parameter \(\alpha = v / c_1\) as the normalized interface speed, which has the meaning of a Mach number. The sign of \(\alpha\) indicates the relative direction of the incident wave with respect to the interface: \(\alpha > 0\) indicates co-propagation, while \(\alpha < 0\) indicates counter-propagation, meaning the incident wave and the moving interface are moving in the same and opposite directions, respectively. Similarly, \(\gamma < 1\) indicates a transition from a faster medium to a slower medium, while \(\gamma > 1\) indicates the opposite. In the acoustic problem considered here, if the velocity difference between media is due to different bulk moduli with equal densities, \(\gamma\) also corresponds to the ratio of characteristic impedances.

\begin{figure}[h!]
    \centering
    \includegraphics[width=0.5\textwidth]{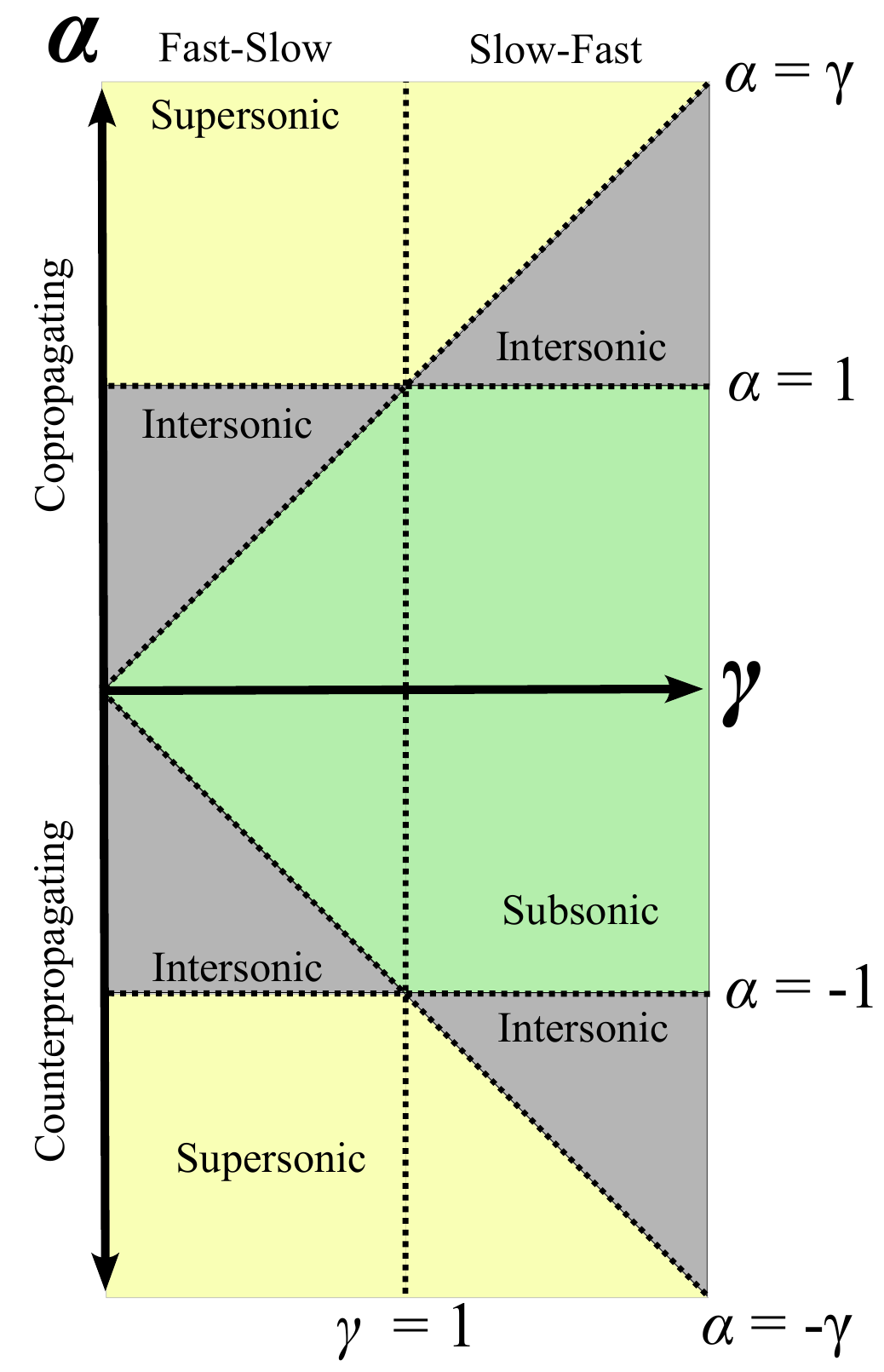}
    \caption{Representation of the different regimes in the $\gamma-\alpha$ diagram.}
    \label{fig:Regime_Sectors}
\end{figure} 
We note that these two parameters are sufficient to classify all interaction regimes. The conditions defining the regimes in the new parameters are: (a) \( |\alpha| < \min(1,\gamma) \) for the subsonic regime, (b) \( |\alpha| < \max(1,\gamma) \) for the supersonic regime, and (c) \( 1 < |\alpha| < \gamma \) or \( \gamma < |\alpha| < 1 \) for the intersonic regime. 

In Fig. \ref{fig:Regime_Sectors}, we present the $(\alpha - \gamma)$ diagram, along with the different interaction regimes, represented in different shading. Specific reference lines for \(\gamma\) and \(\alpha\) mark transitions between regimes: \(\gamma = 1\) corresponds to equal speeds in both media (no interface), and \(\alpha = 0\) corresponds to a stationary interface. Additionally, \(\alpha = 1\) and \(\alpha = -1\) represent an interface moving at the speed of the first medium in the positive (co-propagating) and negative (counter-propagating) directions, respectively. The diagram also includes the lines \(\alpha = \pm \gamma\), where the interface moves at the speed of sound in the second medium. Fig. \ref{fig:Regime_Sectors} illustrates the occurrence of the different regimes in terms of the new parameters. The $(\alpha - \gamma)$ diagram is a valuable tool for visualizing these different interaction regimes and the transitions between them.
Frequency and wavelength changes, as well as scattering coefficients, are readily expressed in terms of the reduced parameters $\alpha$ and \( \gamma \). The results for pressure coefficients are summarized in Table \ref{tab:coefs_pressure}.
\begin{table}[h]
    \begin{center}
        \renewcommand{\arraystretch}{1} 
        \begin{tabular}{|c|c|c|c|}
            \hline
            \textbf{Scattering coefficients} & $\mathcal{F}_p$& $\mathcal{B}_p$& $\mathcal{R}_p$\\
            \hline
            Subsonic & \(\frac{2 \gamma}{1+\gamma}\)& No wave & \(\frac{\gamma -1}{ \gamma +1}\)\\
            \hline
            Supersonic & \(\frac{1+\gamma}{2}\)& \(\frac{1-\gamma}{2}\)& No wave\\
            \hline
            Intersonic ($\gamma>1$) & \( \gamma \frac{|\alpha| + 1}{|\alpha| + \gamma}\)& \( \gamma \frac{|\alpha| - 1}{|\alpha| + \gamma}\)& \(\frac{\gamma-|\alpha|}{\gamma+|\alpha|}\)\\
            \hline
          Intersonic ($\gamma<1$)& \(\frac{\gamma-|\alpha|}{1+|\alpha|}\)& \(\frac{\gamma+|\alpha|}{1+|\alpha|}\)& \(\frac{|\alpha|+1}{|\alpha|-1}\)\\ \hline
          \textbf{Frequency conversion}& $\gamma\frac{\alpha-1}{\alpha-\gamma}$ & $\gamma \frac{\alpha-1}{\alpha+\gamma}$ & $\frac{1-\alpha}{1+\alpha}$\\
          \hline

        \end{tabular}
      \end{center}
    \caption{Pressure scattering coefficients and frequency conversion factors, evaluated for the different interaction regimes of acoustic waves with a moving interface.}

    \label{tab:coefs_pressure}
\end{table}
In the subsonic and supersonic regimes, the scattering coefficients of all waves are independent of \(\alpha\); that is, the scattered wave amplitudes are unaffected by the motion of the interface. In contrast, in the intersonic regime, the coefficients of all waves depend on \(\alpha\), revealing sensitivity to interface speed. This is markedly different from what is observed for electromagnetic waves, which is a consequence of the invariance of Maxwell equations to Lorentz transformations, and the dependence of the field amplitudes of the reference frames \cite{Ostrovskii76,Deck-Leger19a}.

The three--scattered--wave regime occurs when $\gamma < 1$ and $-1<\alpha < 0$ (counter-propagating interface). In all other cases, there is only one scattered wave and a shock wave propagating along the interface, as discussed before.

The expressions for the scattering coefficients in Table \ref{tab:coefs_pressure} match at the boundaries between regimes, as can be verified by substitution. For example, when $\alpha \rightarrow \gamma$, the scattering coefficients in the intersonic regime converge to those in the subsonic regime (for $\gamma > 1$) and the supersonic regime (for $\gamma < 1$), respectively.

\section{Spatio-temporal slabs}
\label{sec:slab}

We now consider a spatio-temporal slab, defined as a region of space bounded by two parallel interfaces separated by a distance $L$, both moving at the same speed (the slab thickness $L$  is then constant). The acoustic properties of the material within the slab differ from those of the surrounding media. The slab generalizes the concept of an acoustic wall (or spatial slab), which is well known in acoustics in the static case \cite{kinsler}. A temporal slab is similarly defined as a sudden change in the properties of a uniform medium over a finite time.

Let \(c_2\) and \(c_1\) denote the sound speeds inside and outside the slab, respectively, and  \(v\) the propagating speed of the slab. The dimensionless parameters ($\gamma, \alpha$) introduced to describe a single interface can also be used to describe wave interactions with the moving slab. Now the presence of two parallel interfaces in the space-time domain introduces additional complexity to the previously discussed scattering problems, since multiple reflections and interferences occur that are not present in the single interface case.

The scattering coefficients for the single interface presented in Table~\ref{tab:coefs_pressure} describe the interactions with either of the two interfaces of the slab, considering that the change in sound velocity (or impedance in general) is reversed at the second interface compared to the first. For example, if the amplitude of the incident wave after interacting with the first interface is determined by the coefficient $\mathcal{F}(\gamma,\alpha)$, then the forward-propagating wave after interacting with the second interface will have its amplitude changed by $\mathcal{F}(1/\gamma,\alpha/\gamma)$. Similarly, the backward-propagating component after interacting with the second interface will have its amplitude determined by $\mathcal{B}(1/\gamma,-\alpha/\gamma)$, because in this case the resulting wave and the interface propagate in opposite directions. The same considerations apply to the reflection coefficients $\mathcal{R}(\gamma,\alpha)$.

The space-time diagrams in Fig. 5 illustrate the individual scattering processes in the subsonic (a) and supersonic (b) regimes, as well as some definitions that will be used later.

\begin{figure}[h!]
    \centering
    \includegraphics[width=0.5\textwidth]{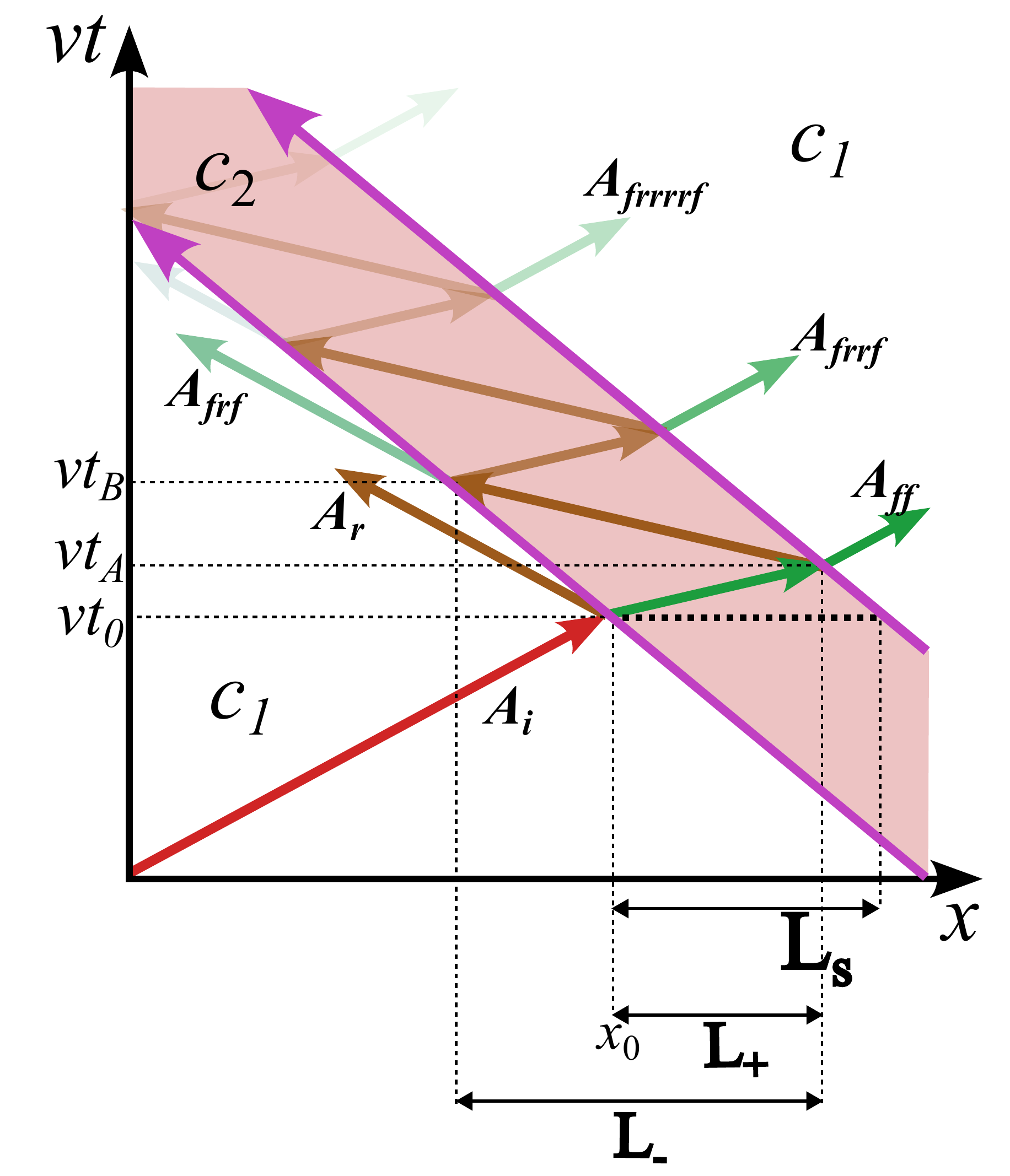}
    
    \caption{Wave interaction processes in a moving slab. Subsonic counter-propagating  regime.}
     \label{fig:slab_sub}
\end{figure}

In the case of a \textbf{subsonic} slab, multiple (in fact, infinite) wave scattering events occur, as shown in Fig. \ref{fig:slab_sub} for the counter-propagating case (the co-propagating case, not shown, is similar). An incident wave with amplitude \( A_i \) (where $A$ represents pressure or particle velocity) experiences subsequent scattering processes which result in wave amplitudes denoted by \( A_{fr^nf} \), where the indices $r$ and $f$ identify the process (reflected or forward waves), and $n$ stands for the order of the process. Excepting the zeroth-order process $A_r$, all wave components experience $n$ internal reflections, traveling the distances $L_+$ and $L_-$ withing the slab.  The waves with odd (even) $n$ add up to generate a wave propagating to the right (left) of the slab. In this superposition, phase differences due to acoustic travel paths  $L_+$ and $L_-$ and frequency changes resulting from interaction with the moving interfaces must be considered, as discussed in the next section.

\begin{figure}[h!]
    \centering
    \includegraphics[width=0.5\textwidth]{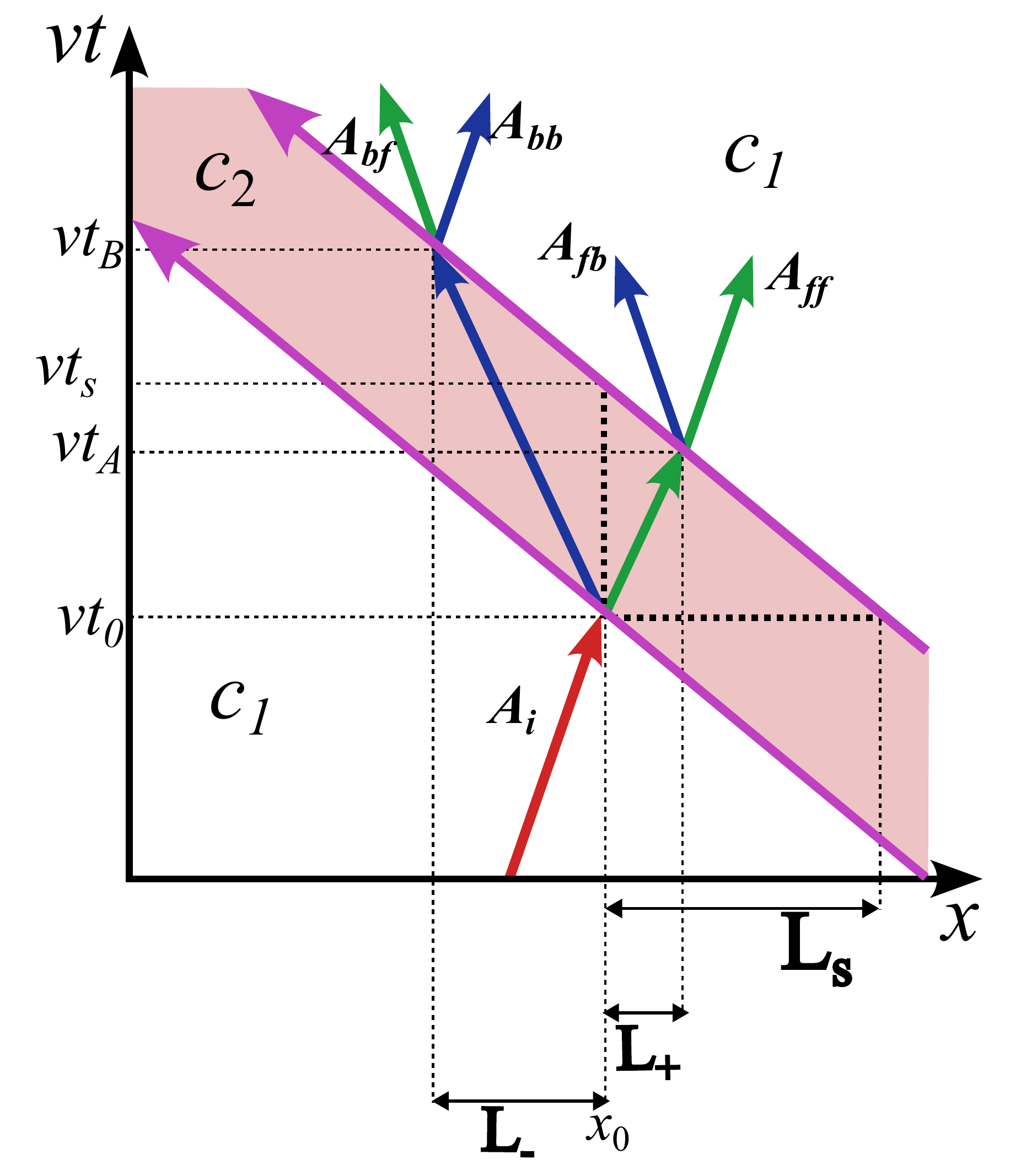}
    \caption{Wave interaction processes in a moving slab. Supersonic counter-propagating regime.}
     \label{fig:slab_sup_count}
    
\end{figure}

In the counter-propagating \textbf{supersonic} regime, as shown in Fig. \ref{fig:slab_sup_count}, the incident wave \( A_i \) interacts with the first interface of the slab and generates two waves that propagate inside the slab: a forward wave \( A_f \) and a backward wave \( A_b \), which propagate over the distances \( L_+ \) and \( L_- \), respectively, in opposite directions. When these waves reach the second interface, each generates a new set of forward and backward waves. A scattered wave is classified as forward or backward depending on whether it propagates in the same or opposite direction as the initial wave \( A_i \). A forward wave generated from a backward wave, such as \( A_{bf} \), is still classified as a backward wave because it continues to propagate in the direction opposite to the original incident wave.
The co-propagating components \( A_{ff} \) and \( A_{bb} \), add up to form the total transmitted wave, while the other two co-propagating components, \( A_{fb} \) and \( A_{bf} \), combine to form the total backward wave. 

\subsection{Frequency conversion in a spatiotemporal slab}

Both frequency and wavelength are modified after interaction with a moving interface by a factor that depends on the interface velocity, as discussed in section \ref{section}. When a wave experiences multiple interactions, the resulting frequency change is determined by the product of all those factors. The final frequency value depends on the type of wave; however, it is independent of the propagation regime (subsonic, supersonic, or intersonic).

We adopt the notation \( \Omega_i=\Omega_i(\gamma,\alpha) \), \( \bar{\Omega}_i=\Omega_i(1/\gamma,\alpha/\gamma) \), and \( \bar{\bar{\Omega}}_i=\Omega_i(1/\gamma,-\alpha/\gamma) \), where $i=\{r, f, b\}$. Using the results from Table \ref{tab:coefs_pressure}, it follows that frequency is conserved for all wave components that are reflected back or transmitted forward by the slab after several round trips. In particular, transmitted components do not experience a frequency change with respect to the incident wave; the change at the first interface is reversed by the second interface, since $\Omega_f \ \bar{\Omega}_f = 1$. Internal reflections within the slab also preserve the frequency, as each round trip satisfies $\bar{\Omega}_r \bar{\bar{\Omega}}_r = 1$. Only the reflected waves exhibit a frequency change, given by \( \Omega_{r} = (1-\alpha)/(1+\alpha) \), which remains the same for all reflected components after $n$ round trips in the slab. Note that the frequency change of the reflected wave is independent of the slab properties, as expected.

\subsection{Scattering coefficients of a spatiotemporal slab}

The amplitudes of the waves scattered by a spatiotemporal slab are determined by the interference of multiple wave components generated by interactions with the slab interfaces. Such interference effects depend on several factors. On one hand, amplitude changes are described by the coefficients discussed for a single interface, as given in Table \ref{tab:coefs_pressure}, which were shown to depend on impedance contrast and interface speed (only in the intersonic regimes). The interference also depends on the phase of the waves, which in turn depends both on the frequency/wavelength changes experienced at the different interactions with the moving interface, as given in Table \ref{tab:coefs_pressure}, and on the distances traveled by the wave, which are also influenced by the motion of the slab. Here, we discuss the calculation of the scattering coefficients for the subsonic and supersonic cases, leaving the discussion of the intersonic case for a future study.

\subsubsection{Subsonic slab}

The reflection coefficient of the slab is defined as the ratio of the total reflected field amplitude to the incident field amplitude. The reflected field is evaluated as the direct reflection at the first interface, $\mathcal{R}^{(0)} = (1-\gamma)/(1+\gamma)$, plus the sum of all waves returning to the original medium after multiple reflections inside the slab. The reflection coefficient of the slab then takes the form
\begin{equation}
   \mathcal{R}(\gamma,\alpha) = \mathcal{R}^{(0)}  +  \sum_{n=1}^\infty \mathcal{R}^{(n)}  e^{i \phi_n},
    \label{eq:Rslab1}
\end{equation}
where amplitude of the $n$-th order reflected wave is given by
\begin{equation}
\mathcal{R}^{(n)}= \mathcal{F}(\gamma) \mathcal{R}(1/\gamma)^{2n-1}\mathcal{F}(1/\gamma)
 \label{eq:Rslab2}
\end{equation}
To evaluate the accumulated phase due to the travel within the slab after $n$ round-trips, $\phi_n$, we consider a reference frame (a listener) moving at the same velocity as the slab. In this case, the interfaces are at rest with respect to the listener, so the measured frequency remains invariant after each reflection for such listener. However, because the slab is moving, the effective distance $L_+$ traveled by the wave in the forward direction differs from the distance $L_{-}$ traveled in the backward direction. These distances can be calculated by noting that $L_\pm = c_2 \Delta t_\pm = L \pm v \Delta t_\pm$, where $\Delta t_\pm$ is the travel time in each directions. Then, the accumulated phase is
$\phi^{(n)}= n  k_2 \left(  L_{+} +L_{-}\right)$, where $L_{+}+L_{-}$ is the total distance traveled by the wave in a round-trip inside the slab, and $k_2=\omega_e/c_2$ is the wavenumber in the slab. Solving for $L_\pm$ we get
 \begin{equation}
      \phi^{(n)}= n  k_2  L\left( \frac{\gamma}{\alpha-\gamma} +\frac{\gamma}{\alpha+\gamma}\right).
      \label{eq:phaseref}
 \end{equation}

 Finally, substituting Eq. (\ref{eq:Rslab2}) and (\ref{eq:phaseref}) into Eq. (\ref{eq:Rslab1}) and summing all the contributions we get

\begin{equation}
 \mathcal{R}_s = \frac{1-\gamma^2}{1+\gamma^2+2 i \gamma \cot{\phi} }
 \label{eq:Rslab3}
\end{equation}
where 
\begin{equation}
 \phi = \frac{\gamma^2}{\gamma^2-\alpha^2}k_2 L
 \label{eq:phasetot}
\end{equation}
\begin{comment}
NOTE: In a stationary reference frame, we must consider frequency changes given by factors in Table  \ref{tab:frequency}, then the expression for the phase reads 
  \begin{equation}
      \phi^{(n)}= n  k_{2}  \Omega_fL_{+}  +  n  k_ {2}   \Omega_f\bar{\Omega}_r L_{-}
      \label{eq:phaseref2}
 \end{equation}
the expression for the reflection coefficient is the same, Eq. (\ref{eq:reflection}), but the phase factor is now
  \begin{equation}
      \phi=  k_2 L \frac{ (1-\alpha) \gamma ^2 \left(\alpha ^2+\gamma
   ^2\right)}{\left(\alpha ^2-\gamma ^2\right)^2}
      \label{eq:phasetot2}
 \end{equation}  
\end{comment}
 
 These expressions apply to both co-propagating and counter-propagating slabs in the subsonic regime. Note that in the specific case of a static slab (\( \alpha = 0 \)), $L_{+} = L_{-} = L$, and the reflection coefficient corresponds to that of a static wall with thickness $L$. 

A similar analysis yields the transmission coefficient as the result of multiple interactions with the slab interfaces:
\begin{equation}
    \mathcal{T}(\gamma,\alpha) = \sum_{n=0}^\infty \mathcal{T}^{(n)}  e^{i \phi_n}
    \label{eq:Tslab1}
\end{equation}
where amplitude of the $n$-th order reflected wave is given by
\begin{equation}
\mathcal{T}^{(n)}= \mathcal{F}(\gamma) \mathcal{R}(1/\gamma)^{2n}\mathcal{F}(1/\gamma),
\label{eq: Tslab2}
\end{equation}
and $\phi_n$ was defined in Eq. (\ref{eq:phaseref}).  Evaluating Eq. (\ref{eq:Tslab1}), the transmission coefficient is the subsonic slab is obtained as
\begin{equation}
 \mathcal{T}_s = \frac{2 i \gamma  e^{-i \phi} \csc{\phi}}{1+\gamma^2+2 i \gamma \cot{\phi} }.
  \label{eq:Tslab3}
  \end{equation}

The intensity (or power) coefficients are obtained from the square of absolute values of the amplitude coefficients. In the subsonic case they obey
\begin{equation}
    |\mathcal{R}_s|^2+|\mathcal{T}_s|^2=1
    \label{eq:energycons}
\end{equation}
where
\begin{equation}
    |\mathcal{R}_s|^2=\frac{\left(\gamma^2-1\right)^2}{\left(\gamma^2+1\right)^2 + 4\gamma^2 \cot^2 \phi}
    \label{eq:intenscoefR}
\end{equation}

In the limit case when $\alpha \rightarrow 0$, Eq. (\ref{eq:phasetot}) reduces to $\phi= k_2L$, and the scattering coefficients  reduce to those of a static interface.  
From Eq.(\ref{eq:intenscoefR}) it also follows that the amplitude of the reflected wave $|\mathcal{R}_s|$ vanishes when $\phi=n\pi$, i.e. when
\begin{equation}
kL=n \pi \left(1 - \frac{\alpha^2}{\gamma^2}   \right) 
\label{eq:interference}
\end{equation}
with $n$ an integer. Equation (\ref{eq:interference}) states that the effective thickness of the slab which results in total transmission is reduced by the motion of the interface (nonzero $\alpha$). 

\subsubsection{Supersonic slab}

For the supersonic case, we use a different approach. Instead of using the slab thickness $L_s$ as a parameter, we use the slab duration, $\Delta t_s$, as shown in Fig. \ref{fig:slab_sup_count}. This approach also includes the case of instantaneous slabs, i.e., slabs propagating at infinite velocity, in the formulation. Otherwise, the relation $L_s = v \Delta t_s$ is used to convert between time and distance representations.

We first calculate the time taken by the forward and backward waves to cross the slab, $\Delta t_{\pm}$. Using the equalities $c_2 \Delta t_- = v (\Delta t_- - \Delta t_s)$ and $c_2 \Delta t_+ = -v (\Delta t_+ -\Delta t_s)$, it follows that, in the reduced notation,
\begin{equation}
\Delta t_{\pm}= \frac{\alpha}{\alpha \pm\gamma} \Delta t_s   
\end{equation}
The amplitude of the forward propagating wave (components $A_{ff}$ and $A_{bb}$ in Fig. \ref{fig:slab_sup_count})
\begin{equation}
\mathcal{F}_s= \mathcal{F}(\gamma) e^{i \omega_e t_+} \mathcal{F}(1/\gamma) + \mathcal{B}(\gamma)e^{i \omega_e t_-} \mathcal{B}(1/\gamma) 
\label{eq:noname1}
\end{equation}
while the backward propagating component (components $A_{fb}$ and $A_{bf}$ in Fig. \ref{fig:slab_sup_count}) reads 
\begin{equation}
\mathcal{B}_s= \mathcal{F}(\gamma) e^{i \omega_e t_+} \mathcal{B}(1/\gamma) + \mathcal{B}(\gamma)e^{i \omega_e t_-} \mathcal{F}(1/\gamma) 
\label{eq:noname2}
\end{equation}

which using the scattering amplitudes in Table \ref{tab:coefs_pressure}  gives the slab scattering coefficients
\begin{equation}
\mathcal{F}_s= \frac{1}{4\gamma}     \left(  (1+\gamma)^2   e^{-i \phi_{-}}  - (1-\gamma)^2   e^{-i \phi_{+}}  \right)    
\label{eq:Fslab4}
\end{equation}
and
\begin{equation}
 \mathcal{B}_s= \frac{1- \gamma}{4\gamma}     \left(  e^{-i \phi_{-}}  -  e^{-i \phi_{+}}  \right)    
 \label{eq:Bslab4}
\end{equation}
where the phases are defined as
\begin{equation}
\phi_{\pm}= \frac{\gamma}{\alpha \pm\gamma} k L  
\label{eq:phasepm}
\end{equation}

The intensity (or power) coefficients are obtained from the square of absolute values of the amplitude coefficients. In the supersonic case they obey
\begin{equation}
    |\mathcal{F}_s|^2-|\mathcal{B}_s|^2=1
    \label{eq:noenergycons}
\end{equation}
where
\begin{equation}
    |\mathcal{B}_s|^2=\frac{\left(\gamma^2-1\right)^2}{4\gamma^2}\sin^2\left( \frac{\gamma^2}{\alpha^2-\gamma^2}k_2 L\right)
    \label{eq:intenscoefB}
\end{equation}

Note that, as in the subsonic case, the amplitude of the backward wave $\mathcal{B}_s$ vanishes when $\phi=n\pi$, leading to the same condition of the subsonic case, Eq. (\ref{eq:interference}), but now as a result of destructive interference between $A_{fb}$ and $A_{bf}$ components in the case of supersonic propagation. 

It also follows from Eqs. (\ref{eq:energycons}), (\ref{eq:intenscoefR}), (\ref{eq:noenergycons}), and (\ref{eq:intenscoefB}) that the value of the intensity coefficients is independent of the sign of $\alpha$, both in the subsonic and supersonic regimes. That is, scattered amplitudes are the same for co- and counter-propagating cases.

\section{Numerical validation}\label{sec:numerical}

\subsection{Implementation of the CIT-FDTD method}

The Finite-Difference Time-Domain (FDTD) method, when applied to solve acoustical problems, is based on discretizing the spatiotemporal domain of interest by defining two staggered grids: one for sound pressure ($p$) and another for particle velocity ($u$), arranged alternately in space and time. This allows discrete equations to be solved using an alternating leapfrog scheme across space and time. However, in the presence of motion (or, more generally, space-time dependence of some parameter) this simulation scheme can exhibit instabilities. To simulate the interaction between acoustic waves and a moving interface within the FDTD framework, we use an effective modeling strategy based on introducing a background velocity field $\mathbf{v}_0$. This background velocity field superimposed on a static interface that is transparent to the flow is equivalent to an interface moving at the same velocity, $\mathbf{v = \mathbf{v}_0}$, as discussed in Sec. \ref{sec:interface}.

The background field is incorporated into the governing equations as a convective term that modifies both Euler’s equation and the continuity equation. Physically, this corresponds to adopting a reference frame in which the medium exhibits effective motion; the wave dynamics are then governed by the fluid equations in the presence of this moving background.

From the perspective of spatiotemporal metamaterials, this approach aligns with interpreting a moving interface as a localized modulation in space and time. The background velocity field $\mathbf{v}_0$ plays the role of a dynamic modulation parameter that modifies the effective acoustic properties. In this way, effects such as frequency conversion and dynamic scattering can be reproduced without altering the computational grid or directly introducing discontinuous moving boundaries. Solving the acoustic equations with an auxiliary background velocity field $\mathbf{v}_0$ not only ensures numerical stability but also provides an accurate physical representation of the wave-interface interaction.

With this analogy, we numerically solve Eqs. (\ref{eq:acoust2}) using a Centered-in-Time Finite-Difference Time-Domain (CIT-FDTD) scheme, as described in \cite{Renterghem07}, where the method was applied to sound propagation in the presence of a background flow. Unlike the standard FDTD approach, CIT-FDTD uses two time steps to update pressure and velocity values, allowing the implementation of a background velocity field. This feature is particularly advantageous for modeling moving interfaces, as it enables the inclusion of a convective background flow in the simulation.
The model equations\cite{Renterghem07} were originally derived for flow problems, and a total velocity field $\mathbf{u}_{tot}=\mathbf{v}_0+\mathbf{u}$, where $\mathbf{u}$ is the (acoustic) particle velocity. The acceleration of a fluid element is expressed as $\partial \mathbf{u}_{tot}/\partial t + \mathbf{u}_{tot} \nabla \cdot \mathbf{u}_{tot}$, representing a Lagrangian description of the flow. Therefore, an additional term $\left(\mathbf{u} \nabla \right) \mathbf{v}_{0}$ appears, which vanishes under the assumption of uniform flow. Linearization for $\mathbf{u}<<\mathbf{v}_0$ finally leads to Eqs. (\ref{eq:acoust2}).
The proposed numerical approach is therefore valid for the problem considered here, where $\mathbf{v}_0$ is assumed to be uniform throughout the entire domain, and the spatial inhomogeneity arises only from the presence of the (static) interface between medium properties.

The discretization of Eqs. (\ref{eq:acoust2}) for implementation of the CIT-FDTD method in the one-dimensional case is performed as follows:
\begin{equation}
p^t = p^{t-2} - 2 \Delta t c^2 \rho_0 \nabla u_x^{t-1} - 2 \Delta t v_0 \cdot \nabla p^{t-1} ,
\label{eq:p-disc}
\end{equation} and
\begin{equation}
u_x^t = u_x^{t-2} - \frac{2 \Delta t}{\Delta h \rho_0} \nabla p^{t-1} - 2 \Delta t v_0 \nabla u_x^{t-1} .
\label{eq:u-disc}
\end{equation}
where \(\Delta t\) is the time step from the temporal discretization of the model, and \(\Delta h\) is the spatial step in the \(x\) direction, as this is a one-dimensional simulation. The variables \(p^t\) and \(u_x^t\) represent the pressure and particle velocity at time \(t\).

For the numerical implementation of the model to study wave interaction with spatio-temporal interfaces, we consider a one-dimensional domain of length \(L = 50 \, \text{m}\) containing two materials with the same density, \(\rho = 1.21 \, \text{kg/m}^3\), but different sound speeds: \(c_1 = 341 \, \text{m/s}\) and \(c_2\), which varies between \(68.2 \, \text{m/s}\) (\(\gamma = 0.2\)) and \(1705 \, \text{m/s}\) (\(\gamma = 5\)). The interfacial speed \(v\) is varied from \(0\) (\(\alpha = 0\)) to \(1364 \, \text{m/s}\) (\(\alpha = 4\)). The system is excited at a boundary by a Ricker wavelet with a central frequency $f = 1000 $ Hz. To minimize reflections within the numerical domain, perfectly matched layers (PMLs) are implemented at both ends. To maintain a consistent Courant number throughout the simulation, the medium is discretized with two different spatial steps, \(\Delta h\), chosen based on the propagation speeds in each medium. A Courant number \(C = 0.25\) is used to ensure the stability of the simulation.

A series of numerical experiments were conducted to validate the CIT-FDTD method applied to the study of spatiotemporal interfaces. In all cases, the speed of sound in the initial medium is kept constant at \(c_1\). For each case, either the speed of sound in the second medium \(c_2\) or the interface velocity \(v\) is fixed while the other parameter is varied. Simulations are also performed for spatiotemporal slabs in both the subsonic and supersonic regimes. The results are presented below.

\subsection{Single interface}

\begin{figure}[h!]
    \centering
    \includegraphics[width=0.7\textwidth]{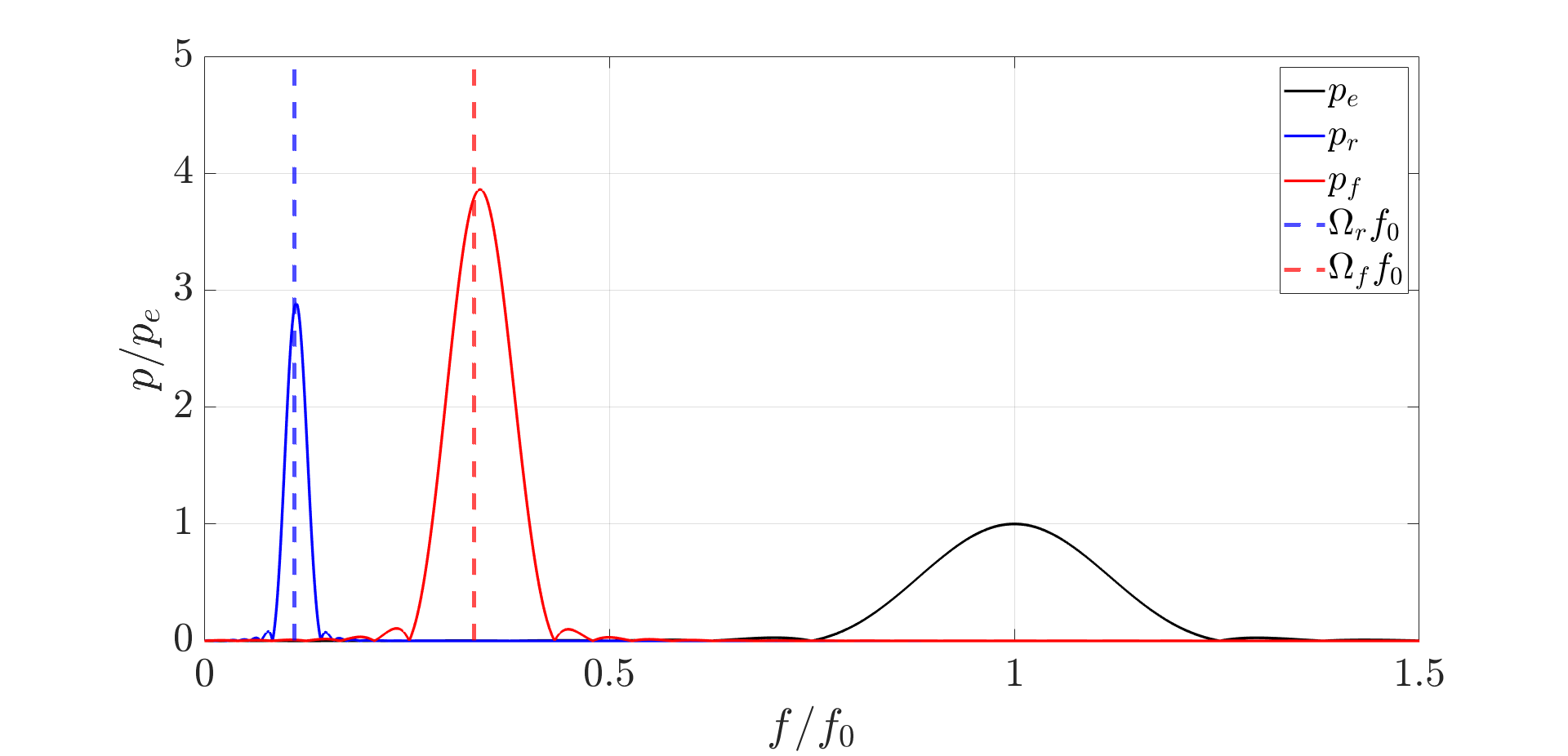}
    \caption{Frequency shift for reflected ($r$) and forward ($f$) waves at a moving interface between two media with a fixed speed contrast $\gamma = 2$ and subsonic interface velocity $\alpha = 0.8$. Curves show results from numerical simulation; vertical dotted lines indicate analytical results, which coincide with the peak frequencies.}
    \label{fig:numeric_frequencyshift}
\end{figure}

In this section, the validity of the analytical expressions obtained for the single interface (frequency shifts and scattering coefficients) is tested by numerically solving Eqs. (\ref{eq:p-disc}) and (\ref{eq:u-disc}) under the previously described conditions. Frequency shift effects are shown in Fig. \ref{fig:numeric_frequencyshift}, where the spectral distributions of field components are obtained for the subsonic, co-propagating case. All signals are broadband, while their peak values correspond to the analytical frequency shift. The same agreement is found in the other cases (not shown).

\begin{figure}[h!]
    \centering
    \includegraphics[width=0.75\textwidth]{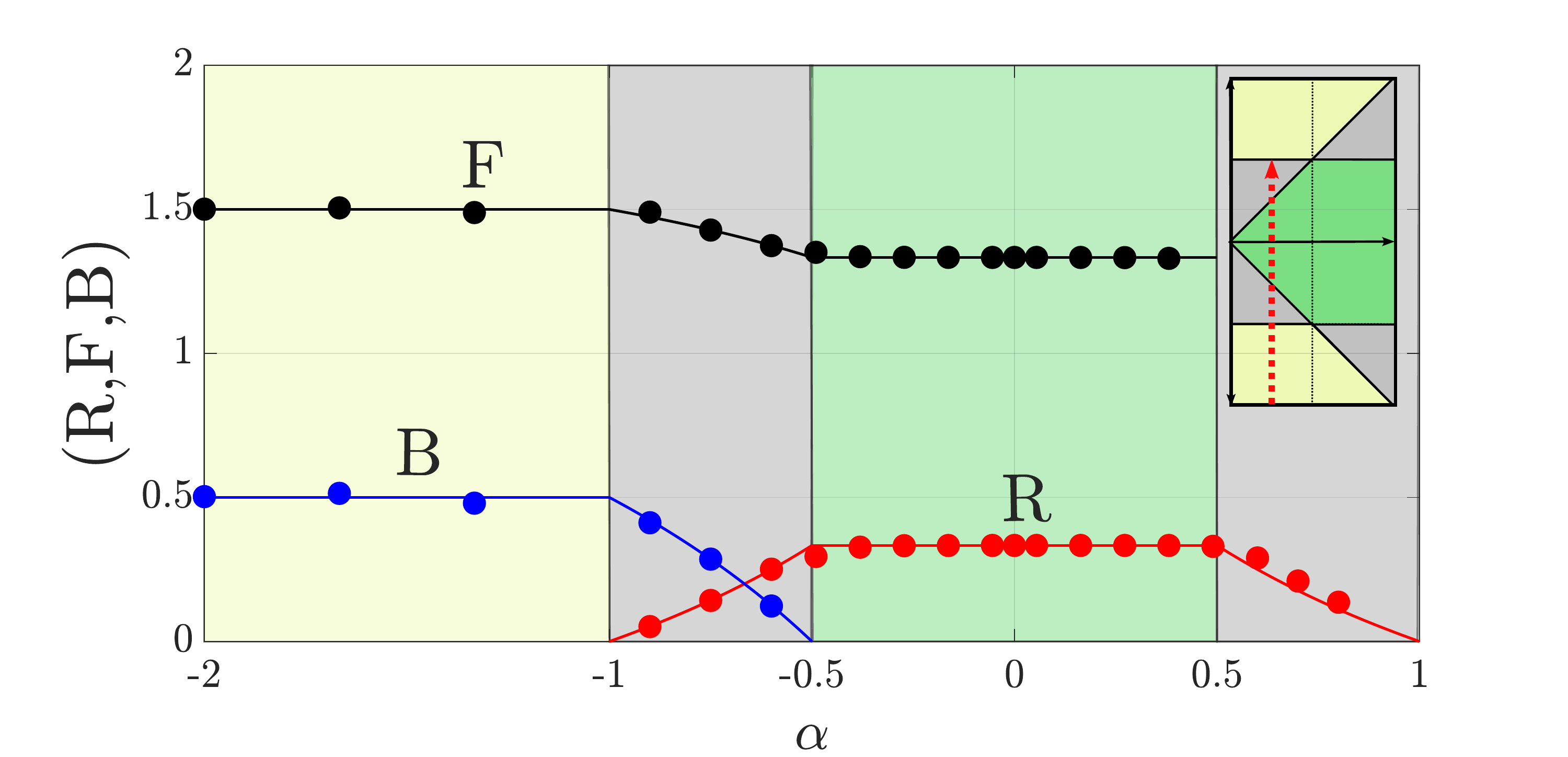}
    \caption{Scattering coefficients for forward ($\mathcal{F}$), backward ($\mathcal{B}$), and reflected ($\mathcal{R}$) waves at a moving interface between two media with a sound speed contrast $\gamma = 0.5$. The normalized interface speed $\alpha$ ranges from $\alpha = -4$ to $\alpha = 1$. Analytical expressions (solid lines) and numerical simulation results (symbols) are shown.}
    \label{fig:gamma_fijo_05}
\end{figure}

The scattering coefficients of the particle velocity for the forward, backward, and reflected waves after their interaction with a moving interface have also been determined. Fig. \ref{fig:gamma_fijo_05} shows the analytical values (see Table \ref{tab:coefs_pressure}), indicated by continuous lines, and the numerical results, indicated by symbols. The specific case of a fast-slow interface with \(\gamma = 0.5\) is shown. The interface speed is treated as a free simulation parameter and is varied in the range \(\alpha = [-4, 1]\). In the supersonic regime (with $|\alpha| < 0.5$, shown in yellow) and the subsonic regime (with $\alpha < 1$, shown in green), the coefficients of the scattered waves are independent of the interface velocity. In contrast, the intersonic regimes (with $-1<\alpha<-0.5$, shown in gray) exhibit a dependence on the interface velocity. In the counter-propagating intersonic regime (with $-1<\alpha<-0.5$), all three waves – reflected, backward, and forward – are present simultaneously. In the co-propagating intersonic regime (at $0.5<\alpha < 1$), however, only the reflected wave is observed. Note the continuity of the coefficients at the transition points.

\begin{figure}[h]
    \centering
    \includegraphics[width=0.65\textwidth]{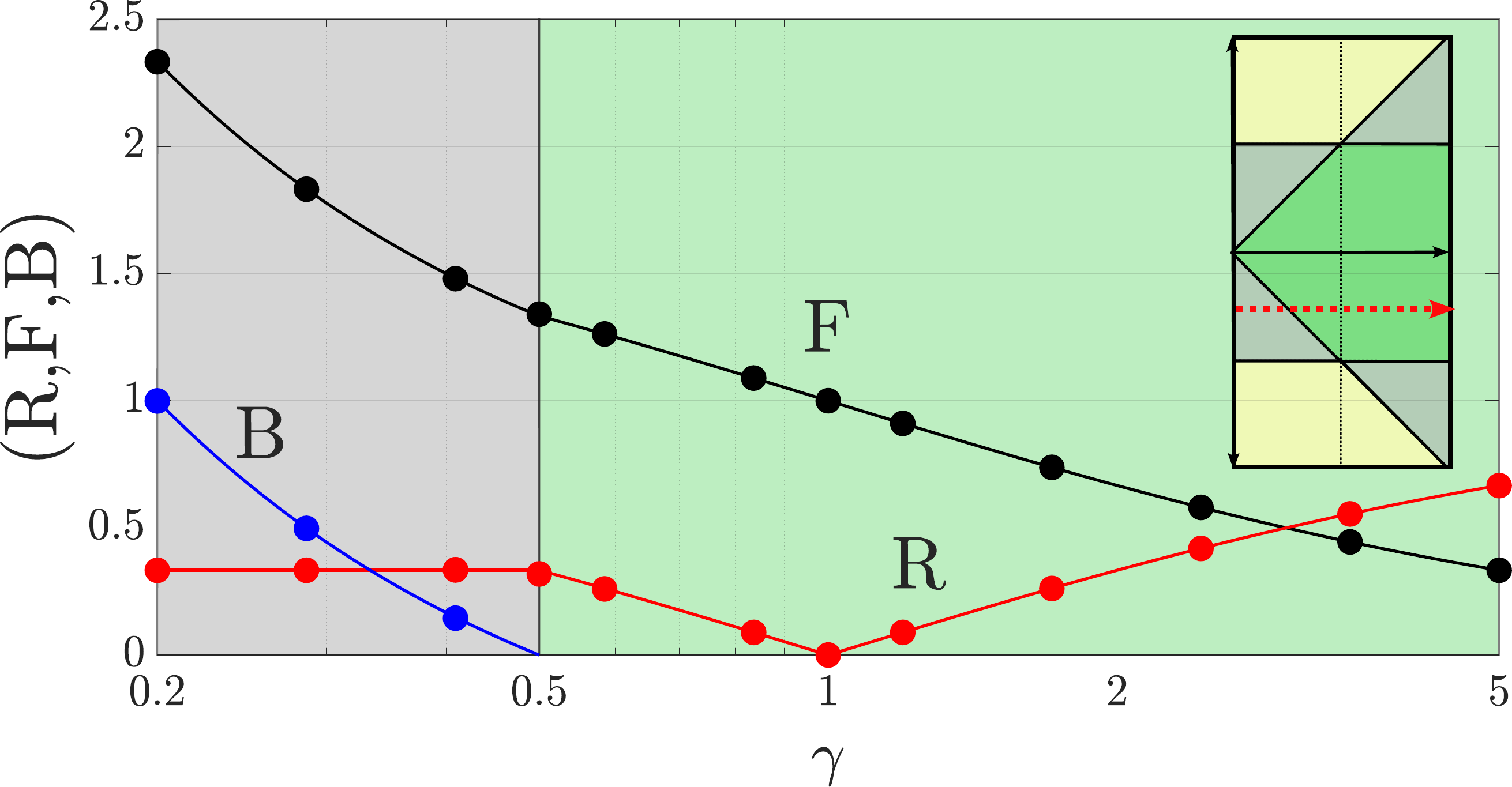}
    \caption{Scattering coefficients of particle velocity for a counter-propagating wave-interface interaction at fixed velocity \(\alpha = -0.5\) and varying \(\gamma\). The path in the \(\alpha\)-\(\gamma\) diagram passes through the intersonic and subsonic regimes. Theoretical results (solid lines) are compared with numerical simulations (symbols).}
    \label{fig:alpha_fijo_05}
\end{figure}

An alternative representation of the scattering coefficients is shown in Fig. \ref{fig:alpha_fijo_05}, where the interface velocity is fixed and the relative sound speed is a free parameter. The counter-propagating case $\alpha=-0.5$ is shown. In the trivial case $\gamma = 1$ (no interface), no backward or reflected wave exists. In the intersonic regime ($0.2<\alpha < 0.5$), all three waves are generated, as expected. The theoretical predictions and numerical simulations show good agreement in this case and in all other cases studied that are not shown here.

\subsection{Spatio-temporal slab}
\begin{figure}[h]
\centering
\begin{minipage}[b]{0.65\textwidth}
    \subfloat{
    \includegraphics[width=\linewidth]{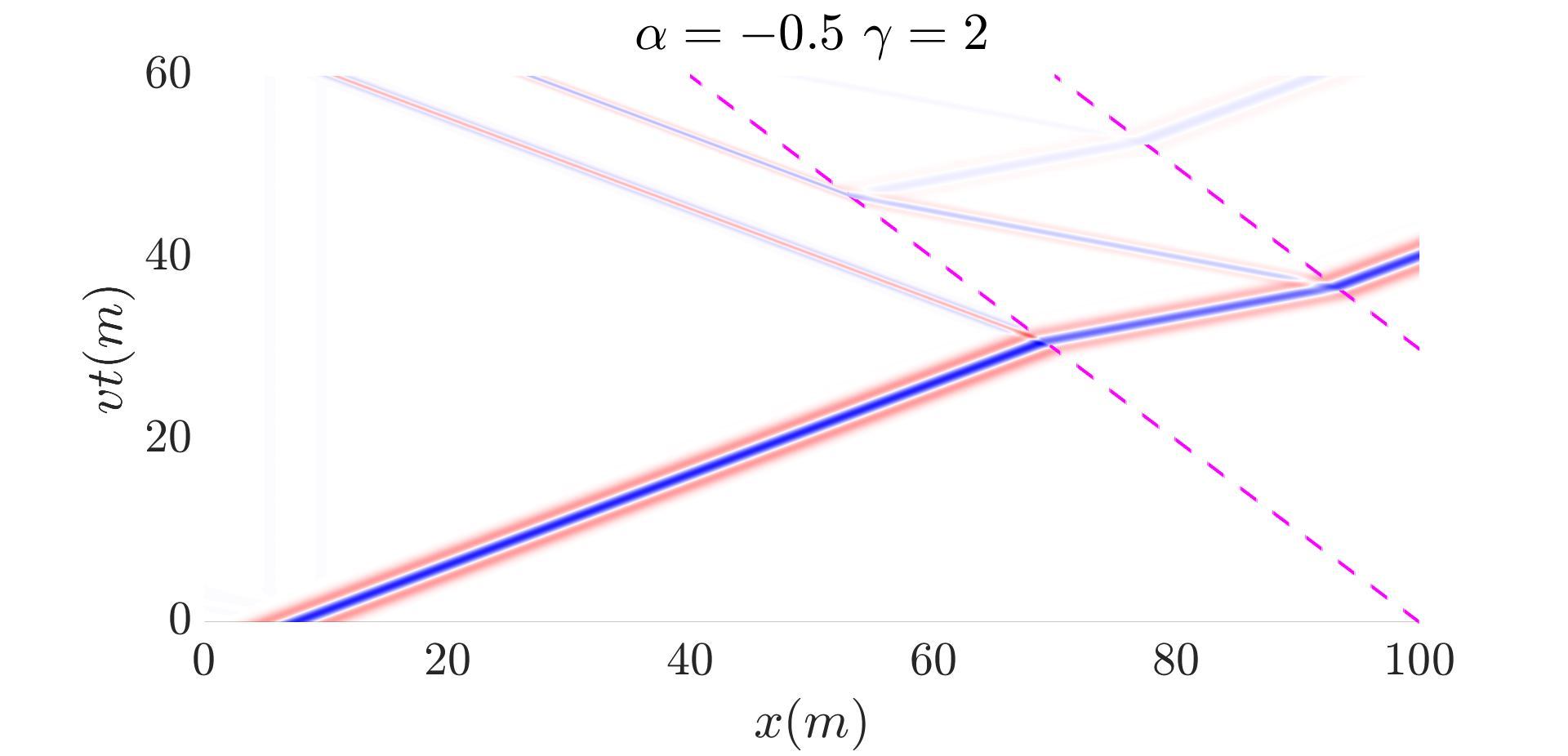}
     }
\end{minipage}
\begin{minipage}[b]{0.65\textwidth}
    \subfloat{
    \includegraphics[width=\linewidth]{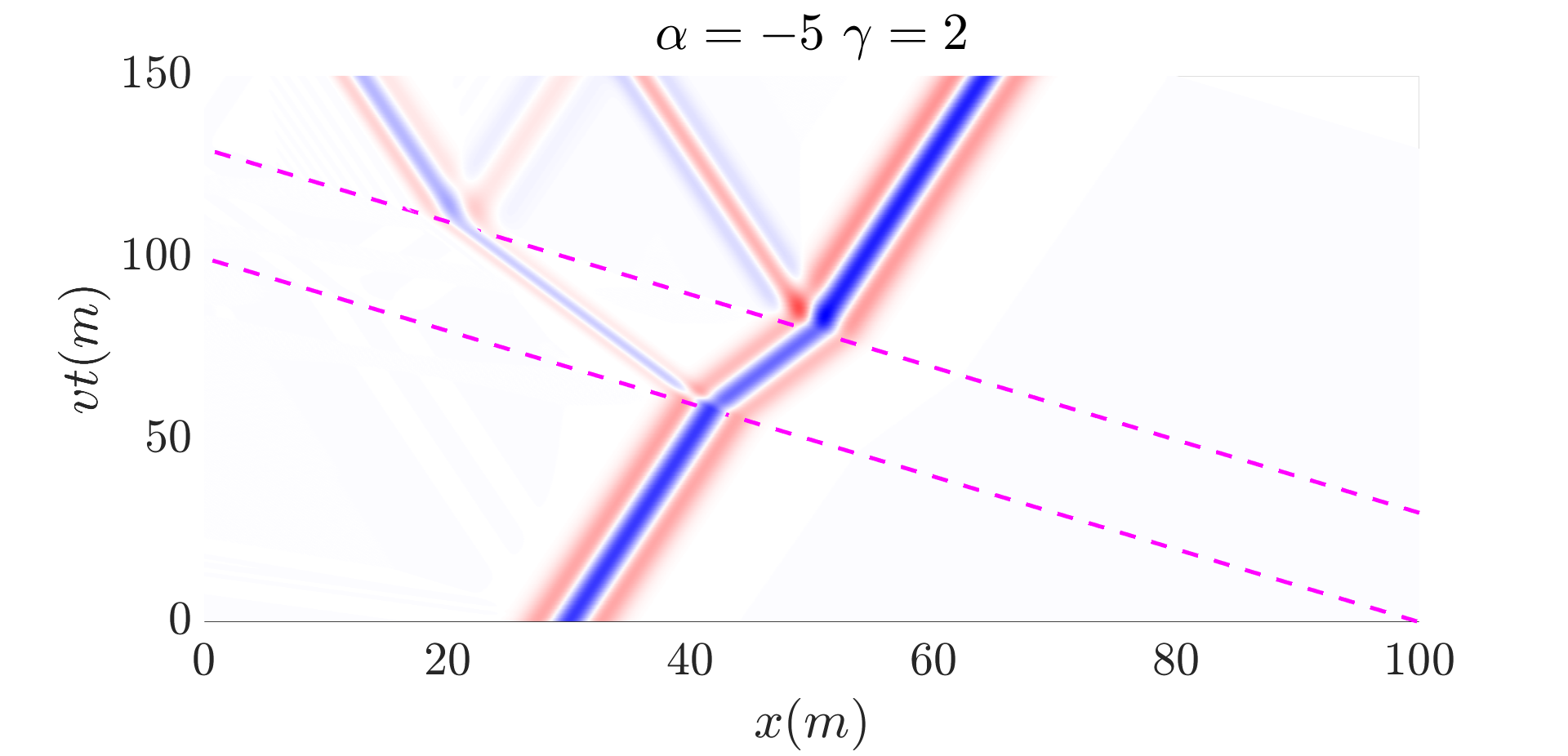}
     }
\end{minipage}

\caption{Space-time representation of a sound pulse passing through a slab moving towards the pulse, as results from FDTD simulation, in subsonic (a) and supersonic (b) cases. The magenta dashed lines indicate the slab interfaces. The $x$-axis shows the length of the entire system, while the $y$-axis shows the distance traveled by the slab, $vt$, where $v$ is the speed of the slab.}
    \label{fig:slab_sim}
\end{figure}

For spatio-temporal slabs, simulations were conducted for the counter-propagating case ($\alpha<0$) in both subsonic and supersonic regimes. Fig. \ref{fig:slab_sim} (a) shows the propagation of a sound pulse and its interactions with a slab in the subsonic regime. The reflected waves, which propagate backward, exhibit space-time compression due to the frequency shift caused by the motion of the slab. The forward waves, however, retain their pulse width, as there is no frequency change after crossing the slab. With each interaction between the waves and the slab interfaces, the amplitude of the scattered components decreases, as discussed previously.

Fig. \ref{fig:slab_sim} (b) shows the propagation in a counter-propagating supersonic slab. As expected, two pairs of forward and backward waves are generated after crossing the slab. As in the subsonic regime, the forward pulses retain their width, while the backward pulses are compressed.

\begin{figure}[h]
\centering
\begin{minipage}[b]{0.65\textwidth}
    \subfloat{
    \includegraphics[width=\linewidth]{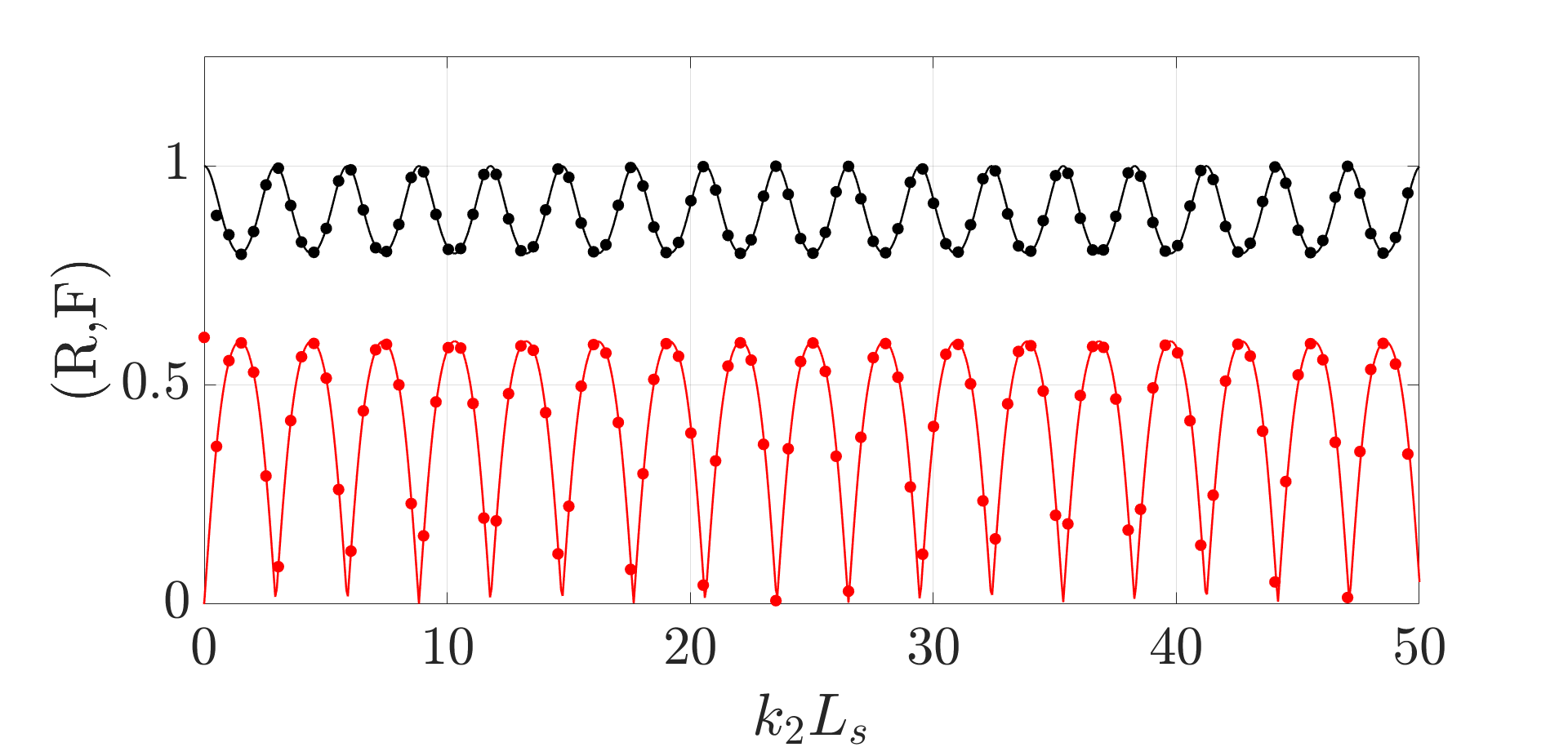}}
\end{minipage}
\begin{minipage}[b]{0.65\textwidth}
    \subfloat{
    \includegraphics[width=\linewidth]{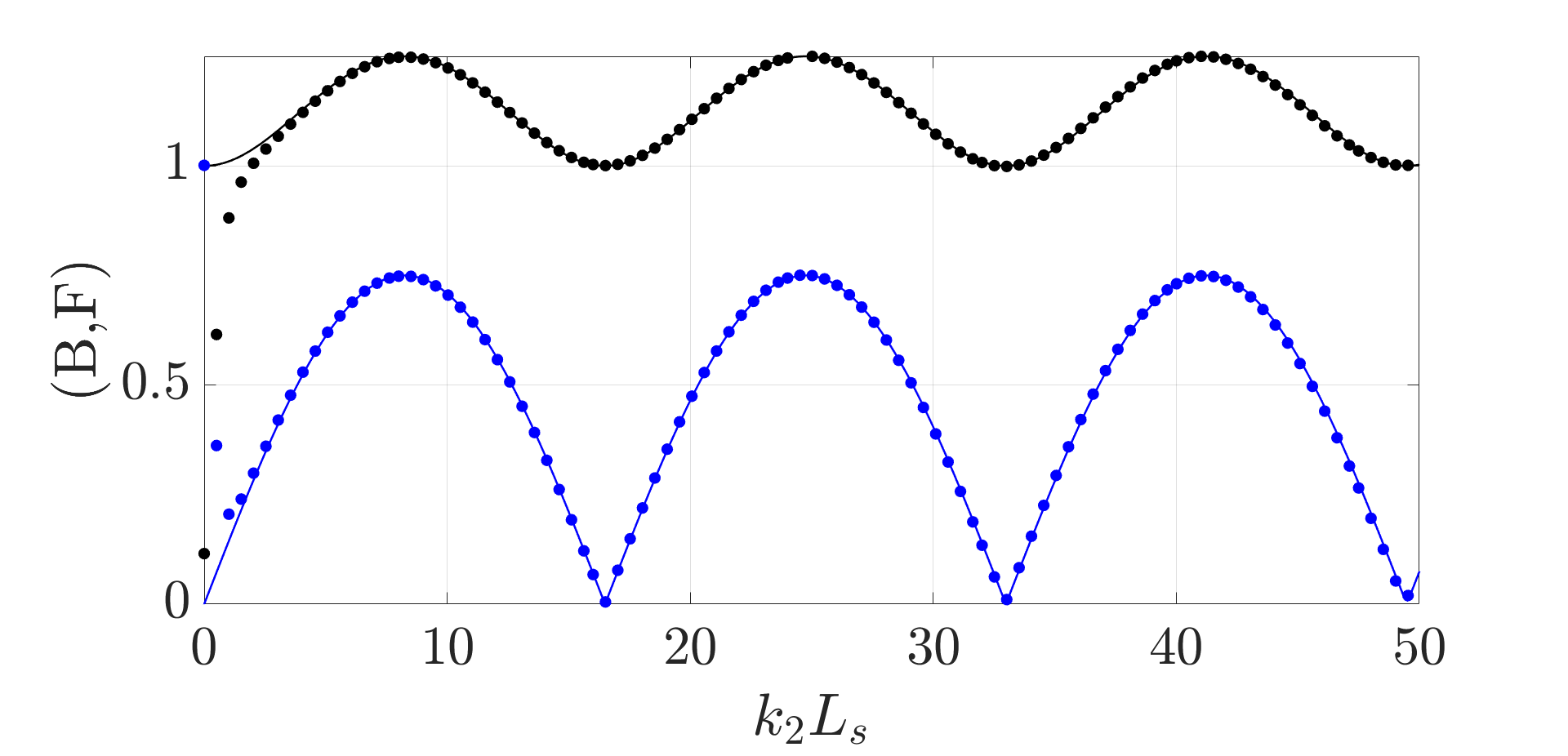}}
\end{minipage}
\caption{Top: scattering coefficients for forward ($F$) and reflected ($R$) waves for a counter-propagating slab with a speed contrast $\gamma = 2$, and Mach number $\alpha=-0.5$.  Solid line: analytical results. Symbols: numerical simulation results. Bottom: scattering coefficients for the forward ($F$) and backward ($B$) waves for a supersonic slab with $\gamma=2$ and Mach number $\alpha=-5$.}
\label{fig:scatslab}
\end{figure}

Finally, the scattering coefficients of the slab, obtained numerically, are shown in Fig. \ref{fig:scatslab}, together with the analytical values derived in the previous section. The absolute values of the scattering coefficients are presented. We note that the agreement is excellent in all cases, validating the numerical method. Since $|\mathcal{R}|$, $|\mathcal{B}|$ and $|\mathcal{F}|$ are independent of the sign of $\alpha$, as follows from Eqs. (\ref{eq:Rslab3}), (\ref{eq:Tslab3}), (\ref{eq:Fslab4}), and (\ref{eq:Bslab4}), only the case of $\alpha > 0$ (co-propagating case) is shown.

\section{Conclusion}

We have presented a study of acoustic wave propagation across spacetime interfaces, under different regimes. The study reveals frequency and amplitude conversion effects with are dependent of the velocity of the interface. The analytical results derived for interfaces and slabs have been tested by numerical modeling.  Simulations solve the 1D first-order acoustic equations using FDTD, and the interface is implemented  as a moving discontinuity in the sound speed. This approach is based on the equivalence between a medium with uniform motion (flowing matter) and a moving parameter interface, separating standing media. Numerical modeling captures frequency and wavenumber conversion effects, as well as scattering amplitudes, with high accuracy. 

The study was performed assuming plane waves under normal incidence at an abrupt interface. Extensions to smooth, multiple, or two-dimensional interfaces may be also performed with the tools described in this work. In this way, the present study can set the basis to  analyze more complex spatiotemporal structures, leading to a better understanding of the acoustic wave behavior in these dynamic materials.
The study also shows other aspects that would require further research.  For example, contrary to what was expected, acoustical and electromagnetic waves behave differently when interacting with a moving interface. Scattered acoustic waves show a strong dependence on interface velocity in the intersonic regime, while they are not dependent on interface velocity in the subsonic/supersonic cases. Electromagnetic waves show the opposite.  
The scattering of waves by a moving boundary has a practical interest in connection with the possibility of using these processes for the transformation of the spectrum, representing an alternative to parametric phenomena, that does not need the use of resonances in the system. These and related issues may guide future studies in this topic.

\section*{Acknowledgement}

This research was developed under grant PID2022-138321NBC22
funded by MICIU/AEI/10.13039/501100011033
and ERDF/EU.

\section*{Bibliography}
%\bibliography{}

\begin{thebibliography}{10}

\bibitem{Keller55}
J.B. Keller, Reflection and Transmission of Sound by a Moving Medium, J. Acoust. Soc. Am. 27, 1044–1047 (1955)

\bibitem{Franken56}
P.A. Franken and U. Ingard, Sound Propagation into a Moving Medium, J. Acoust. Soc. Am. 28, 126–127 (1956)

\bibitem{Miles57}
J.W. Miles, On the Reflection of Sound at an Interface of Relative Motion,  J. Acoust. Soc. Am. 29, 226–228 (1957)

\bibitem{Ribner57}
H.S. Ribner, Reflection, Transmission, and Amplification of Sound by a Moving Medium, J. Acoust. Soc. Am. 29, 435–441 (1957)

\bibitem{Landauer63}
R. Landauer, Velocity Modulation of Propagating Waves, J. App. Phys. 34, 1893-1896 (1963)

\bibitem{Cassedy63}
S. Cassedy and A. A. Oliner, Proc. IEEE 51, 1342 (1963).

\bibitem{Yeh65}
C. Yeh, Reflection and Transmission of Electromagnetic Waves by a Moving Dielectric Medium,  J. App. Phys. 36, 3513-3517 (1965)

\bibitem{Yeh66} 
C. Yeh, K. K. Casei, Reflection and Transmission of Electromagnetic Waves by a Moving Dielectric Slab, Phys. Rev. 144, 665 (1966).

\bibitem{Ramasastry67}
J. Ramasastry and G.Y. Chin,  Wave Interactions with Moving Boundaries, Electronics Lett. 3 (11), 479-481 (1967)

\bibitem{Tsai67}
C. S. Tsai and B. A. Auld, Wave Interactions with Moving Boundaries, J. App. Phys. 38, 2106 (1967)

\bibitem{Landauer73}
R. Landauer and S.T. Peng,  Velocity modulation of propagating waves, J. Appl. Phys. 44,  1156-1161 (1973).

\bibitem{Kong70}
J.A. Kong, Interaction of Acoustic Waves with Moving Media, J. Acoust. Soc. Am. 48, 236–241 (1970)

\bibitem{Steinmetz72}
G.G. Steinmetz and J.J. Singh, Reflection and Transmission of Acoustical Waves from a Layer with Space‐Dependent Velocity 
J. Acoust. Soc. Am. 51, 218–222 (1972)



\bibitem{Ostrovskii67}
Ostrovskii, L. A., Solomin, B.A. Correct formulation of the problem of wave interaction with a moving parameter jump. Radiophysics and Quantum Electronics,10, 666–668, (1967).

\bibitem{Ostrovskii71}
Ostrovskii, L. A., Stepanov, N. S. Nonresonance parametric phenomena in distributed systems. Radiophysics and Quantum Electronics, 14(4), 387–419  (1971).

\bibitem{Ostrovskii76}
Ostrovskii, L. A., Some "moving boundaries paradoxes" in electrodynamics
Sov. Phys.-Usp. 18, 452-458 (1976)

\bibitem{Lurie17}
Lurie, K.A., An introduction to mathematical theory of dynamic materials, 2017, Springer Verlag 

\bibitem{Shui14}
Shui, L. Q., Yue, Z. F., Liu, Y. S., Liu, Q. C., Guo, J. J., One-dimensional linear elastic waves at moving property interface, Wave Motion, 51(7), 1179-1192, (2014).

\bibitem{Delory24}
A. Delory, C. Prada, M. Lanoy, A. Eddi , M. Fink, and F. Lemoult, Elastic Wave Packets Crossing a Space-Time Interface,
Phys. Rev. Lett. 133, 267201 (2024)

\bibitem{Deck-Leger19a}
Z.-L. Deck-Léger, N. Chamanara, M. Skorobogatiy, M.G. Silveirinha,
and C. Caloz, Uniform-velocity spacetime crystals, Advanced Photonics 0560021 (5). (2019).

\bibitem{Deck-Leger19b} Deck-Leger, Z.-L.  Caloz C., 2019, \textit{Scattering at Interluminal Interface},  2019 IEEE International Symposium on Antennas and Propagation and USNC-URSI Radio Science Meeting.

\bibitem{Caloz20a} Caloz C. and Deck-Leger, Z.-L.  \textit{Spacetime Metamaterials - Part I: General Concepts},  IEEE Trans. Antennas and Propag. 68, 1569 - 1582 (2020).

\bibitem{Caloz20b} Caloz C. and Deck-Leger, Z.-L.  \textit{Spacetime Metamaterials - Part II: Theory and applications},  IEEE Trans. Antennas and Propag. 68, 1583 - 1598  (2020).


\bibitem{Deck-leger21}
Deck-Léger, Z. L., Zheng, X., Caloz, C. Photonics 8, 6,202, (2021). 


\bibitem{Bahrami23}
Bahrami, A., Deck-Léger, Z. L., Caloz, C., Electrodynamics of Accelerated-Modulation Space-Time Metamaterials,  Physical Review Applied, 19(5), 054044,  (2023).


\bibitem{kinsler}
L.E. Kinsler, A.B. Frey, A.R. Coppens, F.V. Sanders, Fundamentals of Acoustics, 4th Ed. John Willey and Sons (2000).

\bibitem{Renterghem07}
T. Van Renterghem and Botteldooren, Dick, Applied Acoustics, 68, 201–216, (2007).


\end{thebibliography}

\end{document}